\documentclass[aps,prd,preprint,showpacs,amsmath,amssymb,floatfix]{revtex4}
\usepackage{epsfig}
\usepackage{enumerate}
\usepackage{subfigure}
\usepackage{rotating}
\newlength{\www}

\newcommand{\be}{\begin{equation}}
\newcommand{\ee}{\end{equation}}
\newcommand{\ba}{\begin{eqnarray}}
\newcommand{\ea}{\end{eqnarray}}

\newcommand{\bq}{\begin{equation}}
\newcommand{\eq}{\end{equation}}
\newcommand{\bqa}{\begin{eqnarray}}
\newcommand{\eqa}{\end{eqnarray}}
\newcommand{\ben}{\begin{enumerate}}
\newcommand{\een}{\end{enumerate}}
\newcommand{\bc}{\begin{center}}
\newcommand{\ec}{\end{center}}
\newcommand{\bqb}{\begin{eqnarray*}}
\newcommand{\eqb}{\end{eqnarray*}}

\begin{document}

\title{\vspace{1cm}
Stop-antistop and sbottom-antisbottom production at LHC: a one-loop search for model parameters dependence
}
\author{
M.~Beccaria$^{a,b}$, 
G.~Macorini$^{c, d}$,
L.~Panizzi$^{c, d}$,
F.M.~Renard$^h$ 
and C.~Verzegnassi$^{c, d}$ \\
\vspace{0.4cm}
}

\affiliation{\small
$^a$ $\mbox{Dipartimento di Fisica, Universit\`a del Salento, Italy}$ \\
\vspace{0.2cm}
$^b$ INFN, Sezione di Lecce, Italy\\
\vspace{0.2cm}
$^c$ $\mbox{Dipartimento di Fisica Teorica, Universit\`a di Trieste, Italy}$ \\
\vspace{0.2cm}
$^d$ INFN, Sezione di Trieste, Italy\\
\vspace{0.2cm}
$^h$ Laboratoire de Physique Th\'{e}orique et Astroparticules, 
Universit\'{e} Montpellier II, France
}

\begin{abstract}
We have computed  the  one-loop  electroweak  expression of diagonal and non diagonal stop-antistop and sbottom-antisbottom production from initial state gluons at LHC. We have investigated the possibility that the one-loop effects exhibit a dependence on ``extra'' supersymmetric parameters different from the final squark masses. Our results, given for a choice of twelve SUSY benchmark points in the MSSM with mSUGRA symmetry breaking, show that in some cases a mild dependence might arise, at the percent relative level, of not simple experimental detection.
\end{abstract}


%\begin{flushright}
PTA/07-47\\
%FNT/T 2007/02
%\end{flushright}

\maketitle

\section{Introduction}
\label{sec:intro}

If Supersymmetry exists, and the super particles masses are not all unfairly large, LHC will be able to produce at least a fraction of these new creatures. In particular, the expected values of the cross sections for squark-antisquark pairs production should allow their relatively quick identification. This might be particularly true for the case of final stops, that are supposed to be the lightest squarks in the available theoretical framework of the Minimal Supersymmetric Standard Model (MSSM).
Not surprisingly, three calculations of stop-antistop production already exist, two at the electroweak Born level (including NLO QCD effects) for diagonal~\cite{1} and non diagonal~\cite{2} production and a very recent one~\cite{3} for diagonal production at electroweak NLO. 
For what concerns the dependence on the MSSM involved parameters, at the Born level for diagonal production this is limited to the masses of the two produced stops, conventionally defined
as $\widetilde{t}_1$ (the lighter one) and $\widetilde{t}_2$. For non diagonal production, the calculation of~\cite{2} is done for the electroweak s-channel Born diagram with Z exchange, whose value turns out to be (surprisingly) possibly larger than that of the (kinematically depressed) NLO QCD diagrams, and in principle potentially dependent on the stop mixing angle. Unfortunately, the predicted value of the cross section is in this case very small, and an experimental measurement does not seem to be easily performable, at least in a first LHC running period.
In conclusion, the available calculations at the electroweak Born level for stop-antistop diagonal production only depend on the stop masses,
while those for non diagonal production appear of non trivial experimental determination. Possible observable effects from supersymmetric parameters different from the stop masses might only arise at the next (NLO) electroweak one-loop order.
A complete and exhaustive estimate of  these NLO effects has been performed very recently~\cite{3}, and several important features have been stressed. In particular, the calculation contains, beyond the one-loop corrections to the Born (LHC dominant) initial gluon-gluon state, the 
one-loop corrections to the Born initial quark-antiquark state and also the contributions from the photon-induced gluon-photon fusion channel, with the inclusion of QED effects with soft and hard photon emission. Briefly, one discovers $(a)$ that the one-loop corrections to the less {\em Born relevant} quark-antiquark initial state can be, in some cases, competitive with those coming from the gluon-gluon state and $(b)$ that the effect of the gluon-photon channel can be larger than those of NLO electroweak nature. These results have been derived for an illustrative set of four benchmark points, labelled as SPS1a, SPS1a', SPS2 and SPS5, and are given for the integrated diagonal cross section.
For what concerns the extra SUSY parameter dependence, the analysis of~\cite{3} has been performed for the SPS1a' benchmark point, with the conclusion that it might only arise from the initial gluon-gluon state but, apart from {\em singular} (i.e. of threshold type) effects it would be numerically modest (at the percent level). \\
The process of stop-antistop production is not the only third sfamily case considered in the literature. The production of sbottom-antisbottom, that might be an interesting source e.g. of very light stop subsequent decays, has also been studied in~\cite{2}, together with the stop-antistop one, for both diagonal and non diagonal cases, also considering the mixed stop-sbottom production, at the Born level and for the two benchmark points SPS1a and SPS5. In fact, in~\cite{2} one mSUGRA parameter (the scalar mass $m_0$ or the fermion one $m_{1/2}$) is varied from its 
default benchmark value and the effect on the cross section is shown. In practice, varying these masses changes automatically the stop and sbottom masses, and the plotted changes of the various rates are a pure consequence of the latter variations. For what concerns the non diagonal cases, the considered production mechanism is Z/W exchange at Born level, and the relevant parameters are the stop and sbottom masses, with a possible extra dependence on the stop mixing angle that would deserve a deeper investigation.\\
In conclusion, from the available literature one derives a picture of stop and sbottom production that can be summarized as follows:

\begin{enumerate}
\item 
Diagonal stop-antistop production has been computed at complete NLO electroweak order in the MSSM for mSUGRA symmetry breaking~\cite{3}. A very mild dependence on SUSY parameters different from the stop masses has been found in the considered SPS1a' benchmark point.

\item
Diagonal sbottom-antisbottom production and non diagonal stop and sbottom production have been computed at Born electroweak level~\cite{2}. In both cases the relevant parameters for the total rates are the final squark masses, possibly the stop mixing angle.
\end{enumerate}

\noindent Given these premises, this paper has two different purposes. The first one is that of searching possible cases where the negative conclusions of~\cite{3} might be evaded. With this aim, we have repeated the calculation of the NLO electroweak effects on the gluon initiated  diagonal stop-antistop production, extending the analysis to a 
larger set of benchmark points. Since the ``extra'' (i.e. different from the stop masses) NLO SUSY parameter dependence was excluded in~\cite{3} for the initial quark-antiquark state, we have not included it in our analysis in which we have limited our QED calculations to the derivation of the soft photon contribution. The second purpose is that of performing a NLO electroweak calculation, in search of extra parameter dependence, of some of the processes considered at Born level in~\cite{2}, i.e. non diagonal stop-antistop and diagonal and non diagonal sbottom-antisbottom production (where the relevant masses are now the sbottom ones). Again, we have limited our analysis to the initial gluon-gluon state for the diagonal process and to purely soft photon QED effects. \\
Technically speaking, the paper is organized as follows. In Sections 2 and 3 we give the necessary details of the NLO electroweak calculation, including the treatment of the soft photon contribution, trying to limit the presentation to the essential ingredients. Section 4 contains the definition of the proposed observables and the various computed NLO effects for different choices of SUSY parameters. In Section 5, some tentative conclusions are finally presented.

\section{The kinematics of the processes \lowercase{$g\,g \to \widetilde{t}_a^{\,}\,\widetilde{t}_b^*, ~\widetilde{b}_a^{\,}\,\widetilde{b}_b^*$}}

We discuss in details the case of stop pair production initiated by 2 gluons $g\,g \to \widetilde{t}_a^{\,}\,\widetilde{t}_b^*$.
The kinematic of the process $g\,g \to \widetilde{b}_a^{\,}\,\widetilde{b}_b^*$ is completely analogous to the stop case, 
the only differences being the substitution of stop masses and mixing angles with sbottom ones.

Physical (mixed) stops and antistops are 
denoted as $\widetilde t_a$, and $\widetilde t_b^*$,
with $a,b$ running over 1 and 2. 
 They are obtained from the chirality states 
$\widetilde t_i$, $i=1,2$ standing for L,R as
\bq
\widetilde t_a = R_{ai}\widetilde t_i
\eq
explicitly
\bq
\widetilde t_1=\cos\theta_t\widetilde t_{L}+\sin\theta_t\widetilde t_{R}
~~~~~
\widetilde t_2=-\sin\theta_t\widetilde t_{L}+\cos\theta_t\widetilde t_{R}
\eq 

The momenta, polarization vectors and 
helicities are defined by
\bq
g(p_g, \epsilon(\lambda_g))+g(p'_g, \epsilon'(\lambda'_g))
 \to \widetilde t_a(p_a)+ \widetilde t^*_b(p_b)
\eq
The gluon polarization vectors depend
on the helicities as

\bq
\epsilon(g)=\left(0;{-\lambda_g\over\sqrt{2}},-~{i\over\sqrt{2}},0\right)~~~~~
\epsilon'(g)=\left(0;{\lambda'_g\over\sqrt{2}},-~{i\over\sqrt{2}},0\right)
\label{eps}\eq

We use also the kinematical variables
\bq
s=(p_g+p'_g)^2=(p_a+p_b)^2~~,
u=(p_g-p_b)^2=(p'_g-p_a)^2~~,
t=(p_g-p_a)^2=(p'_g-p_b)^2
\eq

with

\bq
p_g={\sqrt{s}\over2}(1;0,0,1)~~~~~
p'_g={\sqrt{s}\over2}(1;0,0,-1)
\eq

\bq
p_a=(E_a;p\sin\theta,0,p\cos\theta)~~~~~
p_b=(E_b;-p\sin\theta,0,-p\cos\theta)
\eq

\bq
E_a={s+m^2_a-m^2_b\over2\sqrt{s}}~~~~
E_b={s+m^2_b-m^2_a\over2\sqrt{s}}~~~~
p=\sqrt{E^2_a-m^2_a}~~~~\beta={2p\over\sqrt{s}}
\eq
The helicity amplitudes $F_{\lambda_g,\lambda'_g}$,
computed from the Feynman diagrams listed in the next Section
using the polarisation vectors of Eq.~(\ref{eps}),
will appear with various combinations of colours 
of the external particles. 
Firstly, one can write the colour
structure in the form

\bqa
F_{\lambda_g,\lambda'_g}&=&\{~F^1_{\lambda_g,\lambda'_g}
[ if_{ijl}({\lambda^l\over2})]
+F^2_{\lambda_g,\lambda'_g}[{1\over3}\delta_{ij}
+d_{ijl}({\lambda^l\over2})]
\nonumber\\
&&+F^3_{\lambda_g,\lambda'_g}[({\lambda^i\lambda^j\over4})]
+F^4_{\lambda_g,\lambda'_g}[({\lambda^j\lambda^i\over4})]
+F^5[I]~\}_{\alpha\beta}
\label{col}
\eqa
where $i,j$ running from 1 to 8 refer to the gluon colours and 
$\alpha,\beta$ running from 1 to 3 refer to stop and antistop
colours.\\

The polarized cross sections of the process
$g\,g \to \widetilde{t}_a\,\widetilde{t}_b^*$ (averaged over initial
and summed over final colours)
read

\bq
{d\sigma(\lambda_g,\lambda'_g)\over d\cos\theta}=
{\beta\over2048\pi s}
\sum_{col}|F_{\lambda_g,\lambda'_g}|^2
\label{sigpol}\eq
and the unpolarised cross section is
\bq
{d\sigma\over d\cos\theta}=
{1\over4}\sum_{\lambda_g,\lambda'_g}
{d\sigma(\lambda_g,\lambda'_g)\over d\cos\theta}
\eq
The colour summation can be explicitly written
as

\bqa
\sum_{col(ij\alpha\beta)}|F_{\lambda_g,\lambda'_g}|^2&=&
12|F^1_{\lambda_g,\lambda'_g}|^2
+{28\over3}|F^2_{\lambda_g,\lambda'_g}|^2+{16\over3}
(|F^3_{\lambda_g,\lambda'_g}|^2+|F^4_{\lambda_g,\lambda'_g}|^2)\nonumber\\
&&+12(F^1_{\lambda_g,\lambda'_g}F^3_{\lambda_g,\lambda'_g}
-F^1_{\lambda_g,\lambda'_g}F^4_{\lambda_g,\lambda'_g})
+{28\over3}(F^2_{\lambda_g,\lambda'_g}F^3_{\lambda_g,\lambda'_g}
+F^2_{\lambda_g,\lambda'_g}F^4_{\lambda_g,\lambda'_g})\nonumber\\
&&
-~{4\over3}F^3_{\lambda_g,\lambda'_g}
F^4_{\lambda_g,\lambda'_g}
+16F^2_{\lambda_g,\lambda'_g}F^5_{\lambda_g,\lambda'_g}
+8(F^3_{\lambda_g,\lambda'_g}F^5_{\lambda_g,\lambda'_g}
+F^4_{\lambda_g,\lambda'_g}F^5_{\lambda_g,\lambda'_g})\nonumber\\
&&
+24|F^5_{\lambda_g,\lambda'_g}|^2
\label{colsum}\eqa

{\bf The Born terms}

\begin{figure}
\centering
\epsfig{file=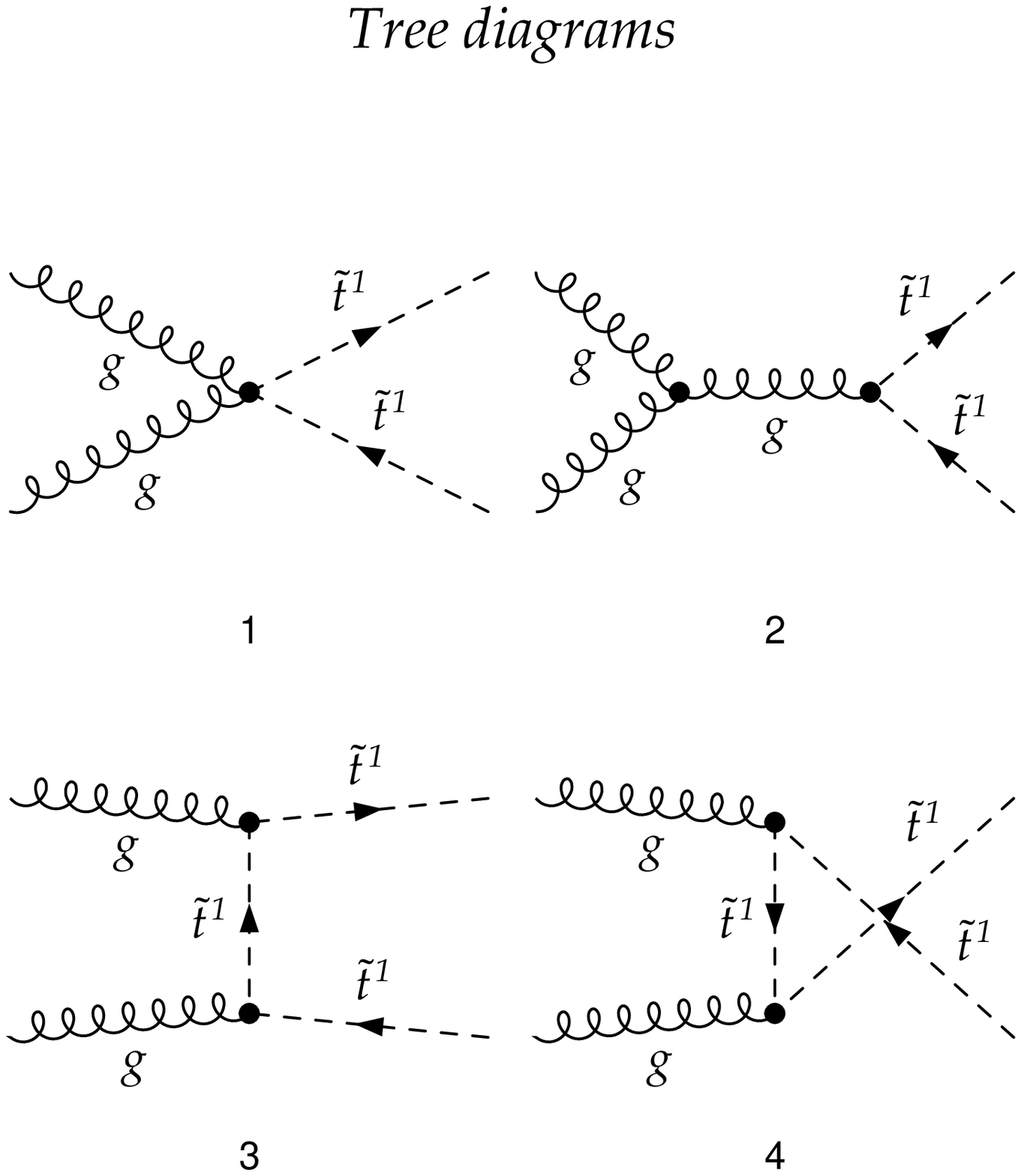, width=14cm, angle=0}
\vspace{1.5cm}
\caption{Tree level diagrams for diagonal production $g\,g\to \widetilde{t}_1\,\widetilde{t}_1^*$.}
\label{fig:fd:tree}
\end{figure}

The Born terms exist only for "diagonal" stop-antistop
pairs ($a\equiv b$).
They are given by 4 diagrams shown in Fig~(\ref{fig:fd:tree}):
\bq
A^{Born}=A^{Born~A}+A^{Born~A'}+A^{Born~B}+A^{Born~C}
\eq 
(A) s-channel gluon exchange:
\bq
A^{Born~A}_{ab}=[if^{ijl}{\lambda^l\over2}] (4\pi \alpha_s)
(\epsilon.\epsilon'){t-u\over s}\delta_{ab}
\eq
(A') 4-leg $g^{i}g^{j}\widetilde t_a\widetilde{t}_a^*$ diagram:
\bq
A^{Born~A'}_{ab}=[{1\over3}\delta_{ij}+d^{ijl}{\lambda^l\over2}]
 (4\pi \alpha_s)(\epsilon.\epsilon')\delta_{ab}
\eq
(B) stop exchange in the t-channel:
\bqa
A^{Born~B}_{ab}&=&-~{16\pi \alpha_s\over
t-m^2_{\widetilde t_a}}~[{\lambda^i\over2}{\lambda^j\over2}]~
(\epsilon.p)(\epsilon'.p)\delta_{ab}
\eqa
(C) stop exchange in the u-channel:
\bqa
A^{Born~C}_{ab}&=&-~{16\pi \alpha_s\over
u-m^2_{\widetilde t_a}}~[{\lambda^j\over2}{\lambda^i\over2}]~
(\epsilon.p)(\epsilon'.p)\delta_{ab}
\eqa
\noindent
(we have used $\epsilon'.p'=-\epsilon'.p$ and
$\epsilon.p'=-\epsilon.p$).

As one sees the Born terms only
involve 2 invariant forms

\bq
I_{1}=(\epsilon.p)(\epsilon'.p)~~~~~~~
I_{2}=(\epsilon.\epsilon')
\label{inv}\eq
(and 4 colour components, $C=1,4$),
so that writing the invariant amplitude as
\bq
A=N_1(s,t,u)I_1+N_2(s,t,u)I_2
\eq
the helicity amplitudes are given by:
\bq
F_{\lambda_g,\lambda'_g}=-~{1\over2}
\lambda_g\lambda'_gp^2\sin^2\theta~ N_1(s,t,u)
+{1\over2}(1+\lambda_g\lambda'_g)~N_2(s,t,u)
\label{helB}\eq
From Eqs.~(\ref{sigpol},\ref{colsum}) and (\ref{helB}) 
one obtains the polarized Born
cross sections
\bq
{d\sigma^{Born}(\lambda_g,\lambda_g)\over d\cos\theta}=
{\pi\alpha^2_s\beta\over24s}(~{m^4_{\widetilde t_a}\over s^2}~)
[~{28+36\beta^2\cos^2\theta\over(1-\beta^2\cos^2\theta)^2}~]
\label{sigGBHV}\eq 
\bq
{d\sigma^{Born}(\lambda_g,-\lambda_g)\over d\cos\theta}=
{\pi\alpha^2_s\beta^5\over384s}
[~{28+36\beta^2\cos^2\theta\over(1-\beta^2\cos^2\theta)^2}~]
\sin^4\theta
\label{sigGBHC}\eq 
in agreement with the results of Ref.\cite{2},\cite{GMW}.
Note that, at this Born level,
$\sigma^{Born}(++)=\sigma^{Born}(--)$ 
and $\sigma^{Born}(+-)=\sigma^{Born}(-+)$. 

\begin{figure}
\centering
\epsfig{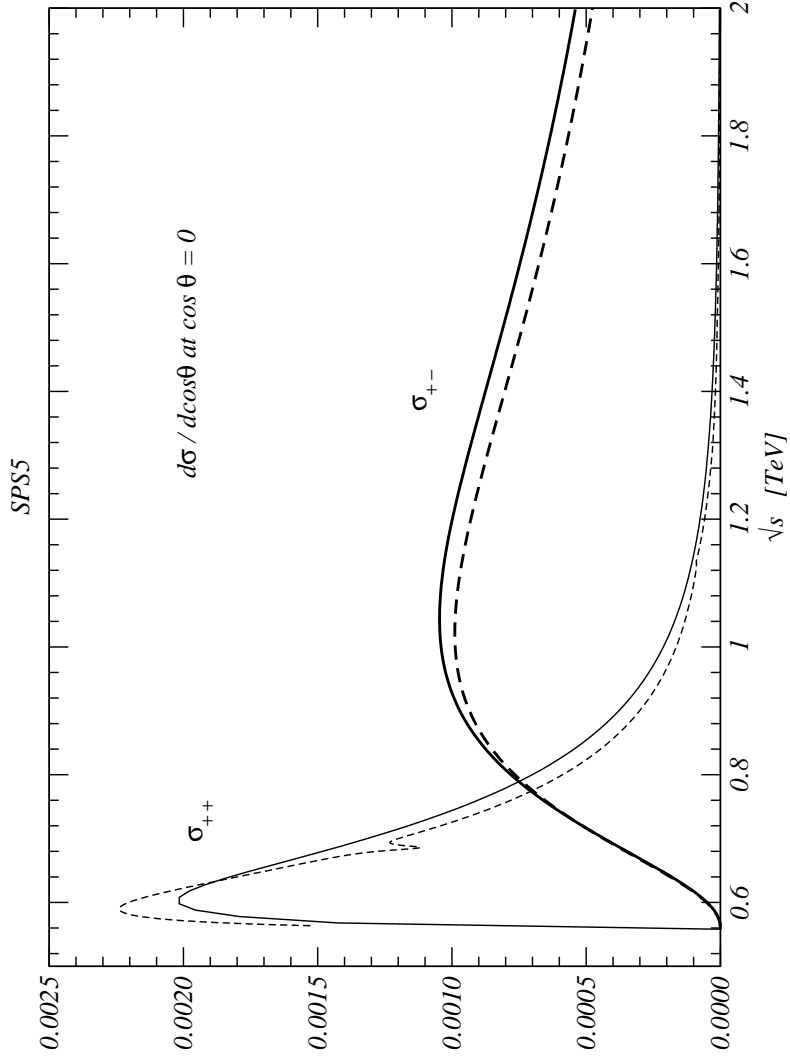}
\vspace{1.5cm}
\caption{Differential cross section at parton level. Energy dependence at fixed angle of the helicity violating (++) and
conserving (+-) components. The dashed lines include the one-loop corrections.}
\label{fig:parton:energy}
\end{figure}

\begin{figure}
\centering
\epsfig{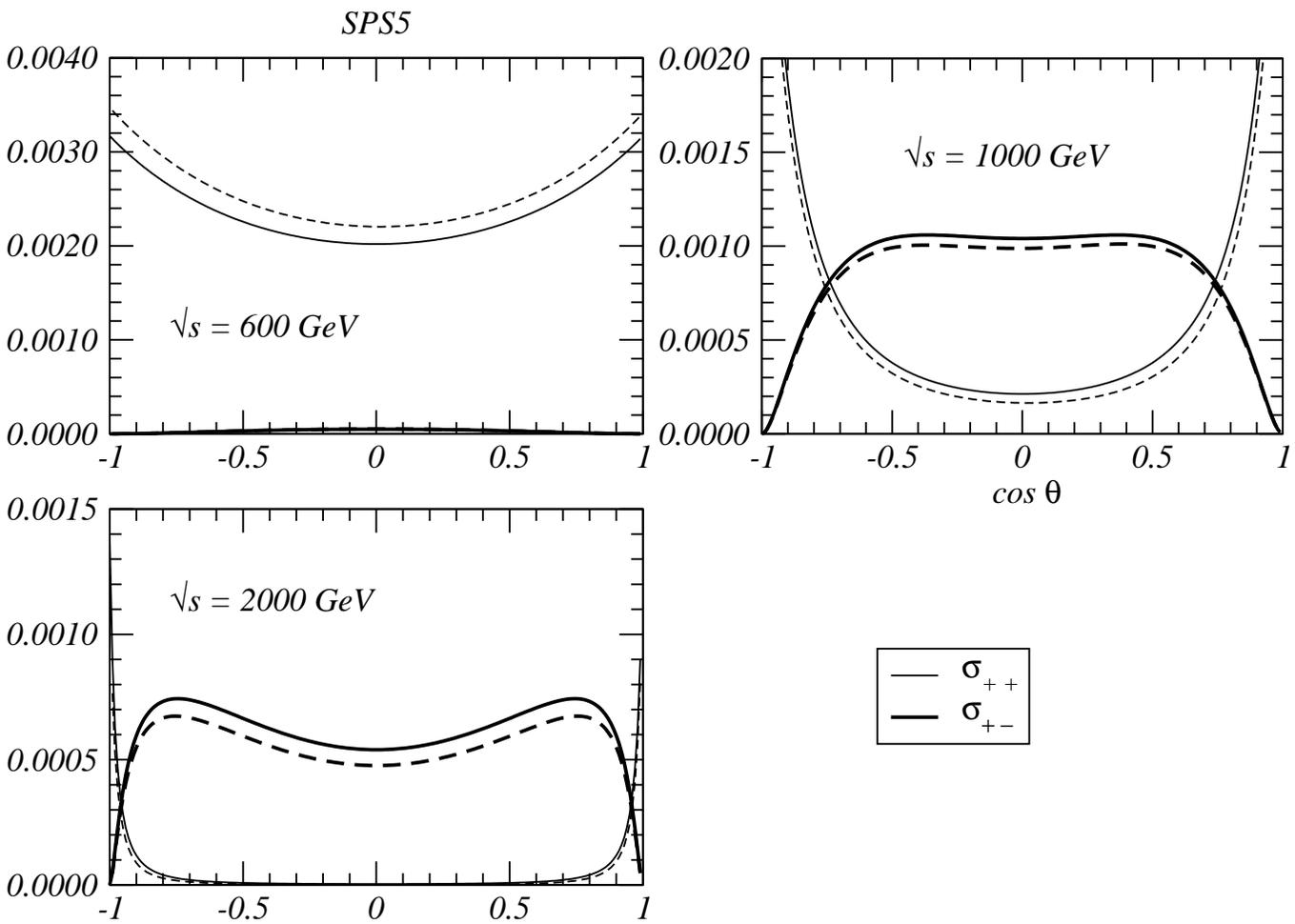}
\vspace{1.5cm}
\caption{Differential cross section at parton level. Angular dependence of the helicity violating (++) and
conserving (+-) components. The dashed lines include the one-loop corrections.}
\label{fig:parton:theta}
\end{figure}

It is useful, for later discussions of one loop effects,
to emphasize the energy and angular dependences of the 
two types of polarized cross sections which are illustrated in Figs.~(\ref{fig:parton:energy},\ref{fig:parton:theta}).

\begin{figure}
\centering
\epsfig{file=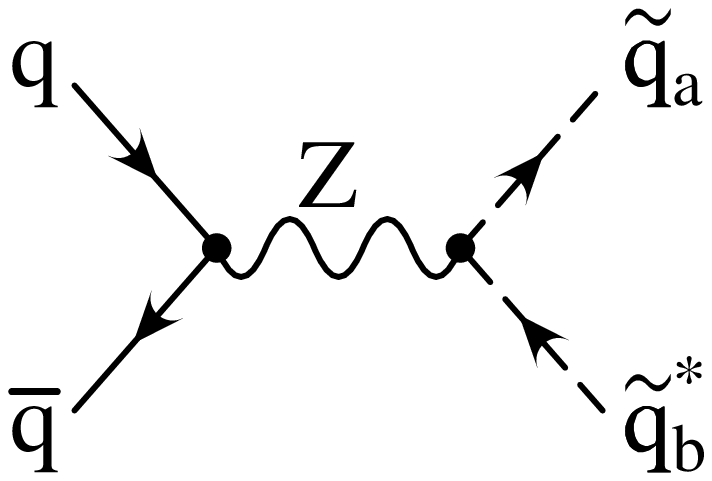, width=6cm, angle=0}
\vspace{1.5cm}
\caption{Born diagram for non-diagonal squark production $q\,\overline{q}\to \widetilde{q}_a\,\widetilde{q}_b^*$, $(a\neq b)$, via $Z$ boson exchange.}
\label{fig:Zexch}
\end{figure}

At low energy the dominant cross sections are the so-called,
\cite{heli}, Gauge Boson Helicity Violating (GBHV) ones
$\sigma(++,--)$ of Eq.~(\ref{sigGBHV}). This arises because the invariant form $I_2$
has no threshold suppression factor, contrarily to $I_1$
which vanishes like $\beta^2$ near threshold. However at
high energy the GBHV cross sections become mass suppressed
like ${m^4_{\widetilde t_a}/s^2}$, as one can check from
Eq.~(\ref{sigGBHV}), in agreement with the general HC rule of Ref.\cite{heli}.
Consequently, as one sees in Fig.~(\ref{fig:parton:energy}), between threshold
($2m_{\widetilde t_a}$) and about $3m_{\widetilde t_a}$, the stop pair
is essentially produced through $\sigma(++,--)$, whereas
for higher energies ($\sqrt{s}>3m_{\widetilde t_a}$) it is
dominated by the GBHC cross sections ($\sigma(+-,-+$) of
Eq.~(\ref{sigGBHC}). In Fig.~(\ref{fig:parton:theta}) we have shown the corresponding angular 
distributions which appear to be also totally different
in the two cases, larger for central angles in $\sigma(+-,-+)$,
see Eq.~(\ref{sigGBHC}), as opposed to forward and backward peaks in 
$\sigma(++,--)$, Eq.~(\ref{sigGBHV}). These various features will be 
essential for understanding the sensitivity to one loop
effects in this process at LHC.\\
As shown in Ref.\cite{3}, the stop pair can also be produced through the $q\bar q$ channel, Fig.~(\ref{fig:Zexch}), with a cross section
\bq
{d\sigma\over d\cos\theta}= {\pi\alpha^2_s\beta^3\over18s}
\sin^2\theta
\eq
and through photon-induced mechanisms present only at NLO. 
Even if at LHC these processes are depressed as compared to the gluon-gluon one because of the smaller (quark) or non-existent (photon) PdF's, the authors of Ref.\cite{3} have shown that the one-loop corrections can be numerically significant or even bigger than those of the gluon fusion initiated process, but essentially independent of extra (i.e. different from the stop mass) SUSY parameters, and for this reason we shall not include them in our analysis.

\section{One loop electroweak corrections to \lowercase{$g\,g\to \widetilde{t}_a^{\,}\,\widetilde{t}_b^*, ~\widetilde{b}_a^{\,}\,\widetilde{b}_b^*$}}

\subsection{Stop pair production: \lowercase{$g\,g\to \widetilde{t}_a^{\,}\,\widetilde{t}_b^*$}}

The one loop electroweak contributions come from 
counter terms (c.t.) and self-energy (s.e.) corrections to the
Born terms, and from triangle and box diagrams. 
We use the on-shell scheme \cite{Hollikscheme}
writing first the c.t.+s.e. corrections as:

For $a=b$:
\bqa
A^{Born+c.t.+s.e.~A}_{aa}&=&A^{Born~A}_{aa}[1+\delta Z_{aa}]
\eqa
\bqa
A^{Born+c.t.+s.e.~A'}_{aa}&=&A^{Born~A'}_{aa}[1+\delta Z_{aa}]
\eqa
\bqa
A^{Born+c.t.+s.e.~B}_{aa}&=&A^{Born~B}_{aa}[1+2\delta Z_{aa}
-{\hat{\Sigma}_{aa}(t)\over t-m^2_{\widetilde t_a}}]
\eqa
\bqa
A^{Born+c.t.+s.e.~C}_{aa}&=&A^{Born~C}_{aa}[1+2\delta Z_{aa}
-{\hat{\Sigma}_{aa}(u)\over u-m^2_{\widetilde t_a}}]
\eqa
and for $a\neq b$, using 
$A^{Born~A,A'}_{aa}=A^{Born~A,A'}_{bb}$
\bqa
A^{Born+c.t.+s.e.~A}_{ab}&=&A^{Born~A}_{aa}
\overline {\delta Z_{ba}}
\eqa
\bqa
A^{Born+c.t.+s.e.~A'}_{ab}&=&A^{Born~A'}_{aa}
\overline {\delta Z_{ba}}
\eqa

\bqa
A^{Born+c.t.+s.e.~B}_{ab}&=&
A^{Born~B}_{aa}[\overline {\delta Z_{ba}}
-{\hat{\Sigma}_{ab}(t)\over 2(t-m^2_{\widetilde t_b})}]
+A^{Born~B}_{bb}[\overline {\delta Z_{ba}}
-{\hat{\Sigma}_{ab}(t)\over 2(t-m^2_{\widetilde t_a})}]
\eqa
\bqa
A^{Born+c.t.+s.e.~C}_{ab}&=&
A^{Born~C}_{aa}[\overline {\delta Z_{ba}}
-{\hat{\Sigma}_{ab}(u)\over 2(u-m^2_{\widetilde t_b})}]
+A^{Born~C}_{bb}[\overline {\delta Z_{ba}}
-{\hat{\Sigma}_{ab}(u)\over 2(u-m^2_{\widetilde t_a})}]
\eqa

with the c.t. terms expressed in terms of stops self-energies

\bq
\delta Z_{aa}=
-[{d\Sigma_{aa}(p^2)\over dp^2}]_{p^2=m^2_{\widetilde t_a}}
\eq

and for $a\neq b$
\bq
\delta Z_{ba}={2\Sigma_{ba}(m^2_{\widetilde t_a})
\over m^2_{\widetilde t_b}-m^2_{\widetilde t_a}}
\eq

\bq
\overline {\delta Z_{ba}}
={1\over2}
[\delta Z^*_{ba}+\delta Z_{ab}]
\eq

the renormalized s.e. functions being given by
\bq
\hat{\Sigma}_{aa}(p^2)=\Sigma_{aa}(p^2)
-\Sigma_{aa}(m^2_{\widetilde t_a})
-(p^2-m^2_{\widetilde t_a})
[{d\Sigma_{aa}(p^2)\over dp^2}]_{p^2=m^2_{\widetilde t_a}}
\eq
and for $a\neq b$
\bq
\hat{\Sigma}_{ba}(p^2)=\Sigma_{ba}(p^2)
+{p^2-m^2_{\widetilde t_b}\over m^2_{\widetilde t_b}-m^2_{\widetilde t_a}}
\Sigma_{ba}(m^2_{\widetilde t_a})
+{p^2-m^2_{\widetilde t_a}\over m^2_{\widetilde t_a}-m^2_{\widetilde t_b}}
\Sigma^*_{ab}(m^2_{\widetilde t_b})
\eq

\begin{figure}
\centering
\epsfig{file=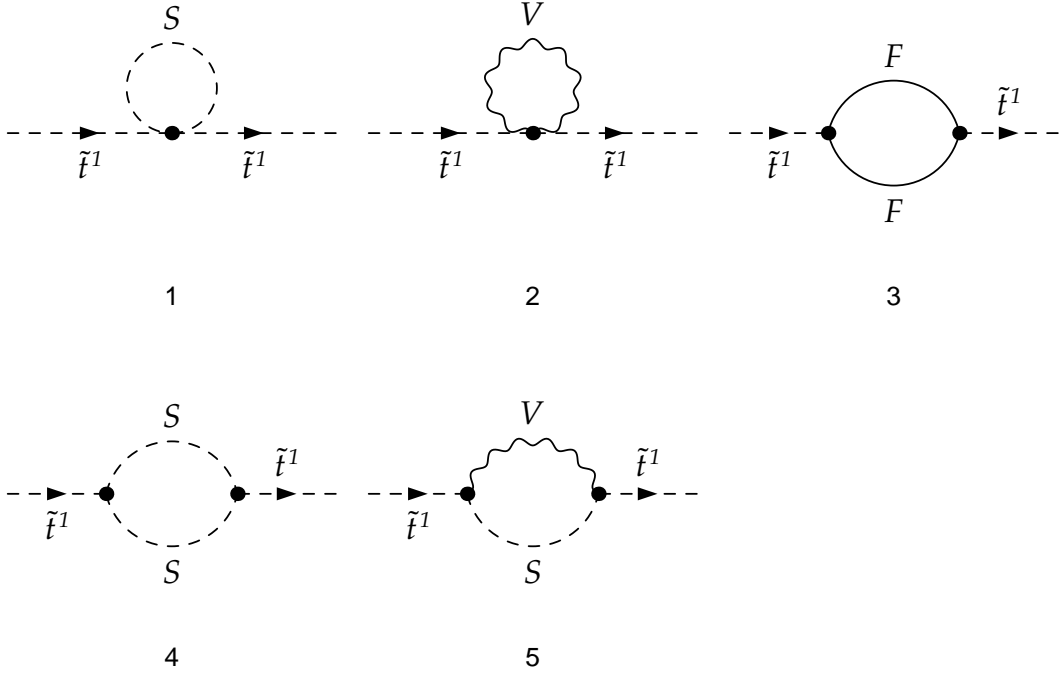, width=14cm, angle=0}
\vspace{1.5cm}
\caption{Self-energy (generic) diagrams for diagonal production $g\,g\to \widetilde{t}_1\,\widetilde{t}_1^*$. They are composed of: scalar and vector tadpoles where the particles can be higgs bosons, sleptons and squarks (1) and SU(2)$\times$U(1) gauge bosons (2); scalar, fermion and scalar-vector bubbles where the particles can be quark-$\chi$ (3), squark-higgs (4) and squark-e.w. gauge boson (5).}
\label{fig:fd:se}
\end{figure}

The needed $\Sigma(p^2)$ functions are obtained from the
various ($\widetilde q V$), ($\widetilde q H$), ($q\chi$) bubbles and
from the gauge boson (V) and the 4-leg ($SS\widetilde t \widetilde t$)
tadpoles depicted in Fig.~(\ref{fig:fd:se}).

\begin{figure}
\centering
\epsfig{file=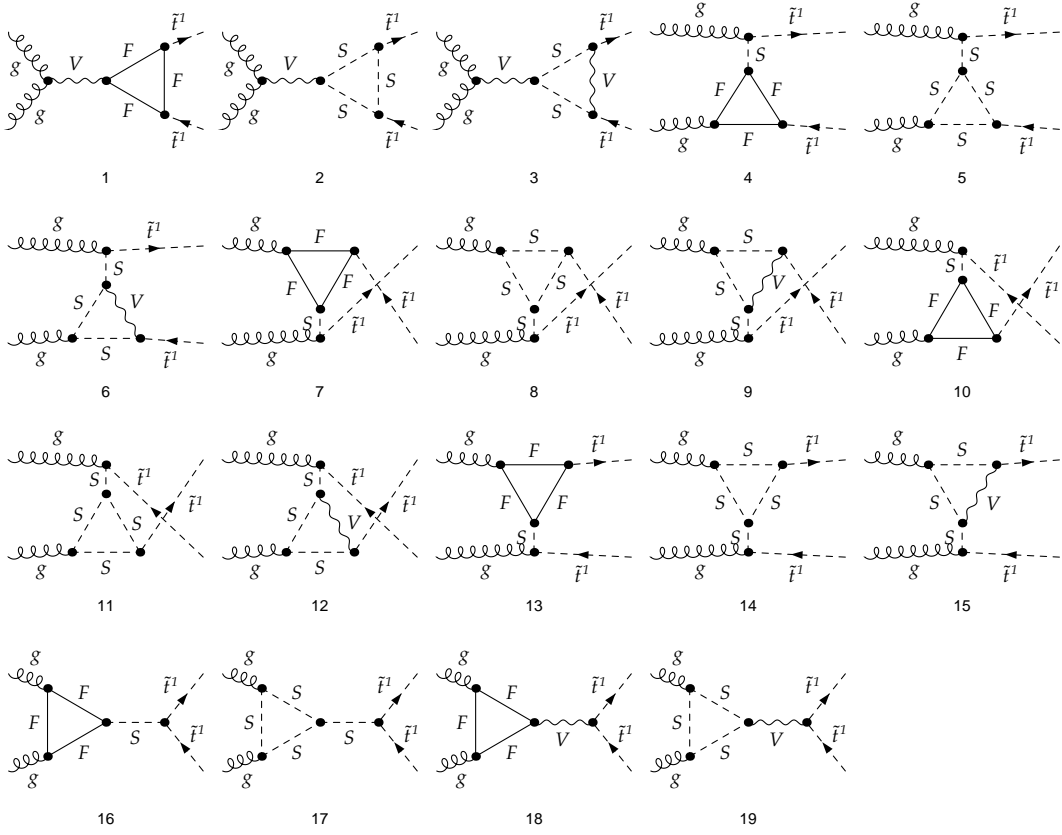, width=14cm, angle=0}
\vspace{1.5cm}
\caption{Up, down, left and right triangle (generic) diagrams for diagonal production $g\,g\to \widetilde{t}_1\,\widetilde{t}_1^*$. In the s-channel diagrams (1), (2), (3) we have labelled the internal gluon as a generic vector, while all the other vector particles are intended to be SU(2)$\times$U(1) gauge bosons. Fermion loops and scalar loops involve quarks-$\chi$ and squarks-higgs respectively, with the exception of diagrams (16), (17), (18) and (19) where the loops involve only quarks and squarks.}
\label{fig:fd:tri1}
\end{figure}

\begin{figure}
\centering
\epsfig{file=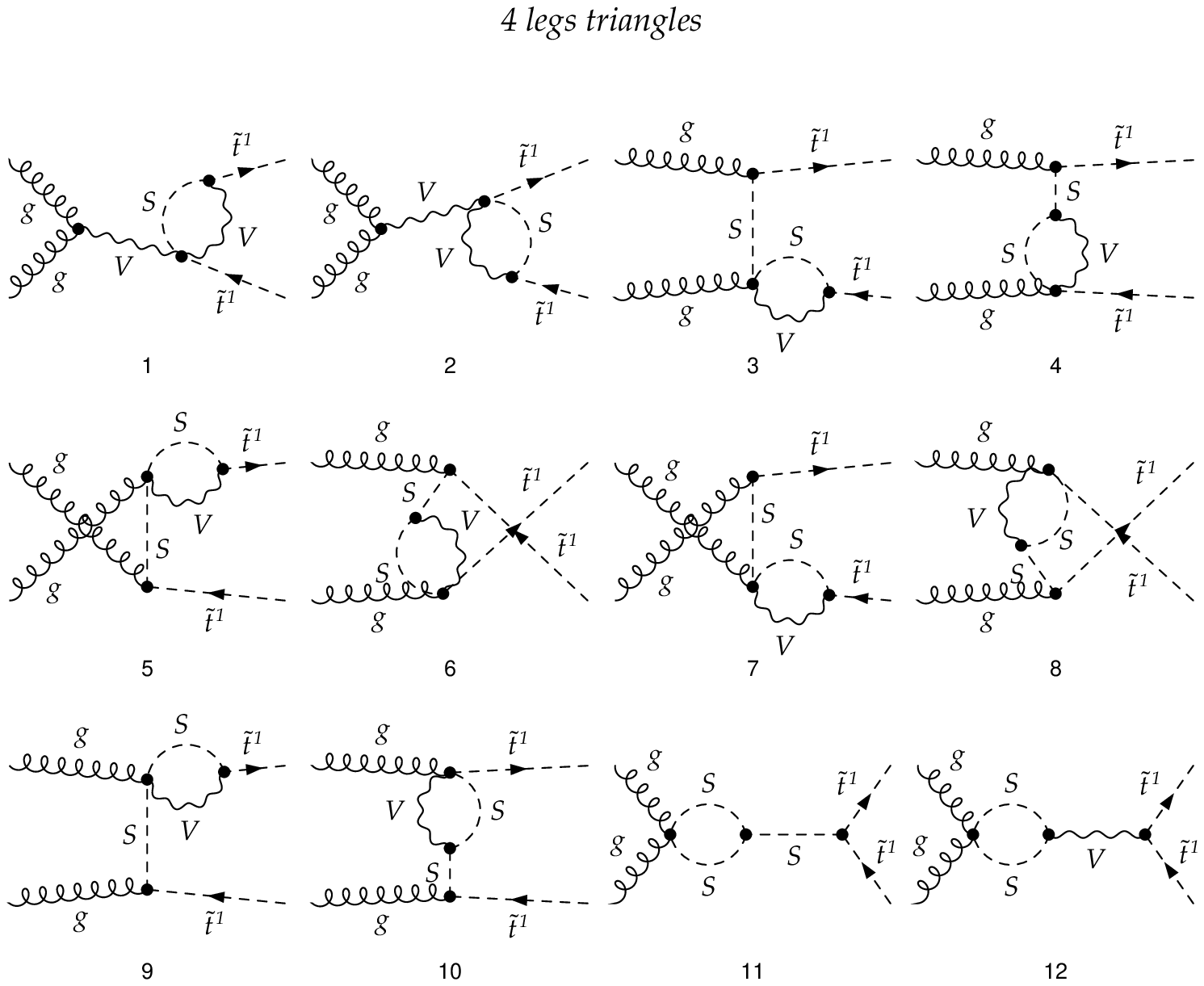, width=14cm, angle=0}
\vspace{1.5cm}
\caption{Four legs triangle (generic) diagrams for diagonal production $g\,g\to \widetilde{t}_1\,\widetilde{t}_1^*$. As in the previous figure we label s-channel internal gluons in diagrams (1) and (2) as vectors, while all the other vectors are e.w. gauge bosons. The s-channel scalar in diagram (11) can be any neutral higgs.}
\label{fig:fd:tri2}
\end{figure}

\begin{figure}
\centering
\epsfig{file=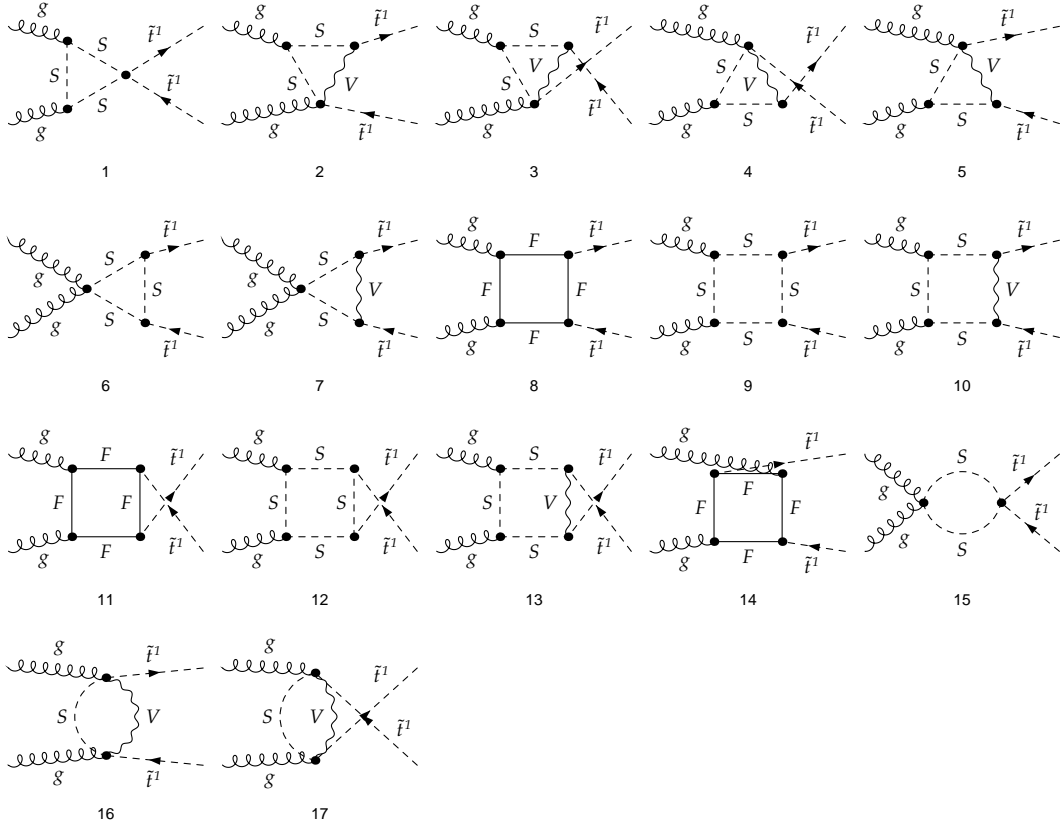, width=14cm, angle=0}
\vspace{1.5cm}
\caption{Box (generic) diagrams for diagonal production $g\,g\to \widetilde{t}_1\,\widetilde{t}_1^*$. Every vector is an e.w. gauge boson; 4-fermions boxes are made of 3 quarks and a $\chi$; 4-scalars boxes are made of 3 squarks and a higgs boson.}
\label{fig:fd:box}
\end{figure}

Triangle and boxes  corrections are shown in Figs.~(\ref{fig:fd:tri1},\ref{fig:fd:tri2},\ref{fig:fd:box}).
They affect respectively
each sector (A), (A'), (B) and (C) 
appearing in the Born case. In the s-channel one
finds "left" and "right" triangles and in the t- and u- 
channels one has "up" and "down" ones. 
Contributions of sector (C)
are obtained from those of sector (B) by symmetrization
rules for the 2 gluons: interchange of momenta, polarization
vectors and colours ($p_g,\epsilon(\lambda_g),i$) and
($p'_g,\epsilon'(\lambda'_g),j$).
The 3 types of boxes can be identified through their
(clockwise) internal contents ($SSVS$), 
($qq\chi q$) and ($SSHS$) for sector (B), the above 
symmetrization rules giving the crossed sector (C);
S refer to all possible scalar states.\\

These electroweak corrections can also be classified into:

\begin{enumerate}
\item[--] {\em gauge} terms due to internal exchanges of gauge bosons
($V=\gamma, Z, W$) and of charginos, neutralinos (through
their gaugino components),
\item[--] {\em Yukawa} terms due to exchanges of Higgs bosons ($H$), and also
charginos, neutralinos  (now through
their higgsino components).
\end{enumerate}

The contributions of these various diagrams
to the helicity amplitudes are obtained after colour 
decomposition according to Eq.~(\ref{col}) and are expressed in
terms of Passariono-Veltman (PV) functions. 
The numerical computation
is then done with a dedicated c++ code exploiting the LoopTools library~\cite{Hahn:1998yk}.

A first check of the computation is obtained by observing
the cancellation of the divergences appearing in
counter terms, self-energies, triangles and
boxes. For some parts
these cancellations occur separately in each sector,
but for other parts they involve contributions from
several sectors as required by gauge invariance.\\

Another type of check is provided by the high
energy behaviour of the helicity amplitudes which
has to satisfy a number of "asymptotic" rules.\\

As already noticed in Sect.II, at high energy, 
neglecting masses the only surviving
Born helicity amplitudes obtained from the addition
of (A+A'+B+C) terms are the GBHC ones:

\bq
F^{Born}_{\lambda_g,~-\lambda_g}=
(4\pi\alpha_s)({\sin^2\theta\over2})
[{c_{ij}\over1-\cos\theta}+{c'_{ij}\over1+\cos\theta}]
\eq
with 
\bq
c_{ij}={1\over3}\delta^{ij}+
d^{ijl}({\lambda^l\over2})+if^{ijl}({\lambda^l\over2})
~~~~
c'_{ij}={1\over3}\delta^{ij}+
d^{ijl}({\lambda^l\over2})-if^{ijl}({\lambda^l\over2})
\eq
in agreement with the theorem given in \cite{heli}, 
whereas the GBHV ones (with  $\lambda_g=\lambda'_g$)
are mass suppressed (vanish like $m^2/s$).\\

From the general logarithmic rules established 
in \cite{MSSMrules} ,
one expects the one loop virtual electroweak contributions
to give, for final unmixed $L,R$ states (before applying the
mixing matrices $R_{ai}$), the following corrections 
to the GBHC Born amplitudes:

\bq
\label{sudakov1}
F_{\lambda_g,~-\lambda_g}=F^{Born}_{\lambda_g,~-\lambda_g}
[1+c_{\widetilde{t}\widetilde{t}}]
\eq

\bq
\label{sudakov2}
c_{\widetilde{t}_L\widetilde{t}_L}=
{\alpha(1+26c^2_W)\over144\pi c^2_Ws^2_W}[2ln\frac{s}{M^2}-ln^2\frac{s}{M_W^2}]
-~{\alpha(\widetilde{m}^2_t+\widetilde{m}^2_b)\over8\pi s^2_WM^2_W}
[ln\frac{s}{M^2}]
\eq
\bq
\label{sudakov3}
c_{\widetilde{t}_R\widetilde{t}_R}=
{\alpha \over9\pi c^2_W}[2ln\frac{s}{M^2}-ln^2\frac{s}{M_W^2}]
-~{\alpha \widetilde{m}^2_t\over4\pi s^2_WM^2_W}[ln\frac{s}{M^2}]
\eq
\bq
\label{sudakov4}
\widetilde{m}_t={m_t\over\sin\beta}~~~~~~
\widetilde{m}_b={m_b\over\cos\beta}
\eq

in which one identifies the "gauge" and the "Yukawa"
parts. M is a typical mass scale whose precise value
does not matter at Log accuracy.\\

We have checked analytically (by taking the leading logarithmic
expressions of the PV functions)
that the various self-energy, triangle and box
contributions reproduce 
the above expressions in both gauge and Yukawa sectors.\\

We conclude this Section by briefly discussing the treatment of infrared singularities.
As usual, QED radiation effects can be split into a soft part which is infrared (IR) singular and a 
hard part including the emission of photons with an energy which is not small compared to the process energy scale.
In this paper, we have only included the soft part which is necessary in order to cancel any the IR 
singularities associated with the photonic virtual corrections. As we have outlined in the Introduction, since we are only searching for extra SUSY parameter dependence, we have not included the hard part of QED effects. \newline
We denote by ${\cal A}^{\rm Born}$ and ${\cal A}^{\rm 1\ loop}$ any invariant helicity scattering amplitude
evaluated at Born or one loop level. IR divergences are regulated by a small photon mass $\lambda$.
IR cancellation holds for every helicity channel separately and we checked it numerically by taking the $\lambda\to 0$
limit of our calculation. The real radiation factorizes on the Born amplitude leading to 
\be
\left({\cal A}^{\rm Born}\right)^2 \left(1 + \frac{\alpha}{2\pi}\delta_s\right) + 2 {\cal A}^{\rm Born}\ {\cal A}^{\rm 1\ loop} = \mbox{IR finite}.
\ee
The universal correction factor $\delta_S$ takes into account the emission of soft real 
photons with energy from $\lambda$ up to $\Delta E_\gamma^{\rm max} \ll \sqrt{s}$~\cite{'tHooft:1978xw}.
In our analysis, we have fixed $\Delta E_\gamma^{\rm max} = 0.1$ GeV.

\subsection{Sbottom pair production: \lowercase{$g\,g\to \widetilde{b}_a^{\,}\,\widetilde{b}_b^*$}}

The treatment of the one loop corrections for the sbottom case is again analogous to that of the stop, but the particles involved in the loops are different, so the numerical results for the one loop contributions obtained in the stop production process cannot be trivially transposed to the sbottom case. In practice, all the expression given in the above section are to be ``mirrored'' substituting every top-tagged quantity with its bottom-tagged counterpart. For this reason we give only a brief overview of the main differences that arise between the two processes.

Since the main parameters that controls the processes are the masses of the final state squarks, we start from some considerations about how such masses affect the observables we are going to analyze. As it is possible to see in Tab.~(\ref{tab:bench}), the masses of stop and sbottom squarks change within a wide range of values depending on the scenario considered, and the thresholds for the production of $\widetilde{q}_a\widetilde{q}_b^*$ vary accordingly affecting the values of the cross section. Thus, since at tree-level the only difference is the sbottom masses instead of the stop masses in $t-m^2$ and $u-m^2$, considering scenarios with not too different masses, the tre-level cross sections should be comparable. In any case, at high energy all the masses can be neglected, so the cross sections are identical to a great approximation.
 
At one loop level, two type of differences appear: a) the different masses in the various propagators, b) the different couplings in gauge, SUSY gauge and Yukawa couplings. This second type can be very simply pointed out by
comparing the Sudakov coefficients controlling the
high energy behaviour:

\bq
\label{sudakov5}
c_{\tilde{b}_L\tilde{b}_L}=c_{\tilde{t}_L\tilde{t}_L}=
{\alpha(1+26c^2_W)\over144\pi c^2_Ws^2_W}[2ln\frac{s}{M^2}-ln^2\frac{s}{M_W^2}]
-~{\alpha(\tilde{m}^2_t+\tilde{m}^2_b)\over8\pi s^2_WM^2_W}
[ln\frac{s}{M^2}]
\eq
\bq
\label{sudakov6}
c_{\tilde{t}_R\tilde{t}_R}=
{\alpha \over9\pi c^2_W}[2ln\frac{s}{M^2}-ln^2\frac{s}{M_W^2}]
-~{\alpha(\tilde{m}^2_t)\over4\pi s^2_WM^2_W}[ln\frac{s}{M^2}]
\eq

\bq
\label{sudakov7}
c_{\tilde{b}_R\tilde{b}_R}=
{\alpha \over36\pi c^2_W}[2ln\frac{s}{M^2}-ln^2\frac{s}{M_W^2}]
-~{\alpha(\tilde{m}^2_b)\over4\pi s^2_WM^2_W}[ln\frac{s}{M^2}]
\eq

\noindent the only difference coming from the R part.

Again this should give only a slight difference at high energy when mass effects are negligible.

\section{Results}

\subsection{Stop pair production: \lowercase{$g\,g\to \widetilde{t}_a^{\,}\,\widetilde{t}_b^*$}}

Our starting observable for this process is the invariant mass 
distribution defined as 
\ba
\label{eq:basic}
\frac{d\sigma(pp \stackrel{gg}{\to} \widetilde{t}_a^{\phantom{*}}\,\widetilde{t}_b^*+X)}{dM_{\rm inv}} &=& \int \, 
dx_1 \, dx_2 \, d\cos\theta \,g(x_1, \mu) \, g(x_2, \mu) \nonumber \\ 
&\times & \frac{d\sigma_{gg\to \widetilde{t}_a^{\phantom{*}}\,\widetilde{t}_b^*}}{d\cos\theta}
\, \delta(\sqrt{x_1 x_2 S} - M_{\rm inv}) \, ,
\ea
\noindent
where $\sqrt{S}$ is the proton-proton c.m. energy, $M_{\rm inv}$ is the $\widetilde{t}_a+\widetilde{t}_b^*$ invariant mass, $\theta$ is the stop squark scattering angle in the partonic c.m. frame, and $g(x_i, \mu)$ are the distributions of the gluon inside the proton with a momentum fraction $x_i$ at the scale $\mu$. We have used  the LO PDF set CTEQ6L~\cite{PDF} with $\mu = m_{\widetilde{t}_a}+m_{\widetilde{t}_b}$. As we already mentioned, we include soft QED real radiation in order  to cancel IR singularities. For the $2\to 2 + \gamma (\rm soft)$ process we can identify $M_{\rm inv}$ with the partonic c.m. energy $\sqrt{s}$. The shift induced by hard QCD radiation has been previously estimated for $t\,\overline{t}$ production in~\cite{Beccaria:2004sx} and found to be at the level of a few percents. Since our observables will be defined by integrating over a wide range of $M_{\rm inv}$ values, such a shift will be irrelevant for our conclusions.

For our purposes, we have considered the total rate $\sigma_{\rm tot}$ of the process defined by integrating the distribution 
$d\sigma/dM_{\rm inv}$ over the full range of invariant mass values, from the threshold $m_{{\widetilde t}_a}+m_{{\widetilde t}_b}$, for the diagonal light squark production ($\widetilde{t}_1\widetilde{t}_1^*$) and for the non-diagonal case ($\widetilde{t}_1\widetilde{t}_2^*+\widetilde{t}_2\widetilde{t}_1^*$).

Our analysis has been performed for a choice of a large number of SUSY benchmark points. More specifically we have considered 12 mSUGRA inspired points: the eight SPS points (SPS1a, SPS1a', SPS1a slope, SPS2-6)~\cite{SPS} which allow, as far as SPS1a, SPS1a', SPS1a slope, SPS2 and SPS5 are concerned, a direct comparison with the results of~\cite{2},~\cite{3}, the two SU1, SU6 ATLAS points~\cite{DC2} and  two light SUSY scenarios LS1 and LS2 discussed in~\cite{Beccaria:2006ir}. In Tab.~(\ref{tab:bench}) we have listed the values of the chosen set of parameters: $m_0$, $m_{1/2}$, $A_0$, $\tan\beta$ and sign~$\mu$.

\begin{figure}
\centering
\epsfig{file=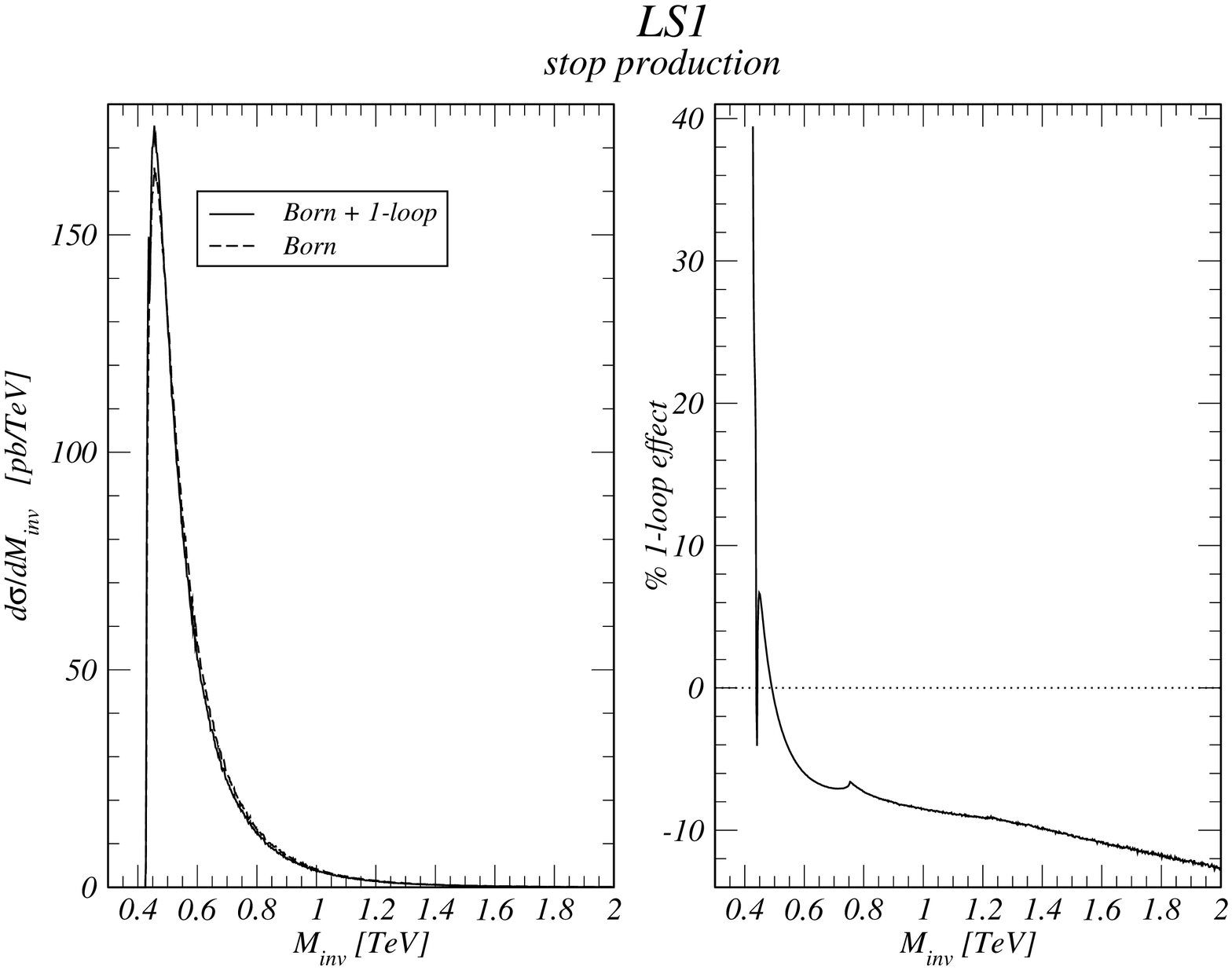, width=18cm, angle=90}
\vspace{1.5cm}
\caption{LS1, Born and one-loop distribution $d\sigma/dM_{\rm inv}$ for stop production. The right panel shows the percentual relative effect.}
\label{fig:distrib:LS1st}
\end{figure}

\begin{figure}
\centering
\epsfig{file=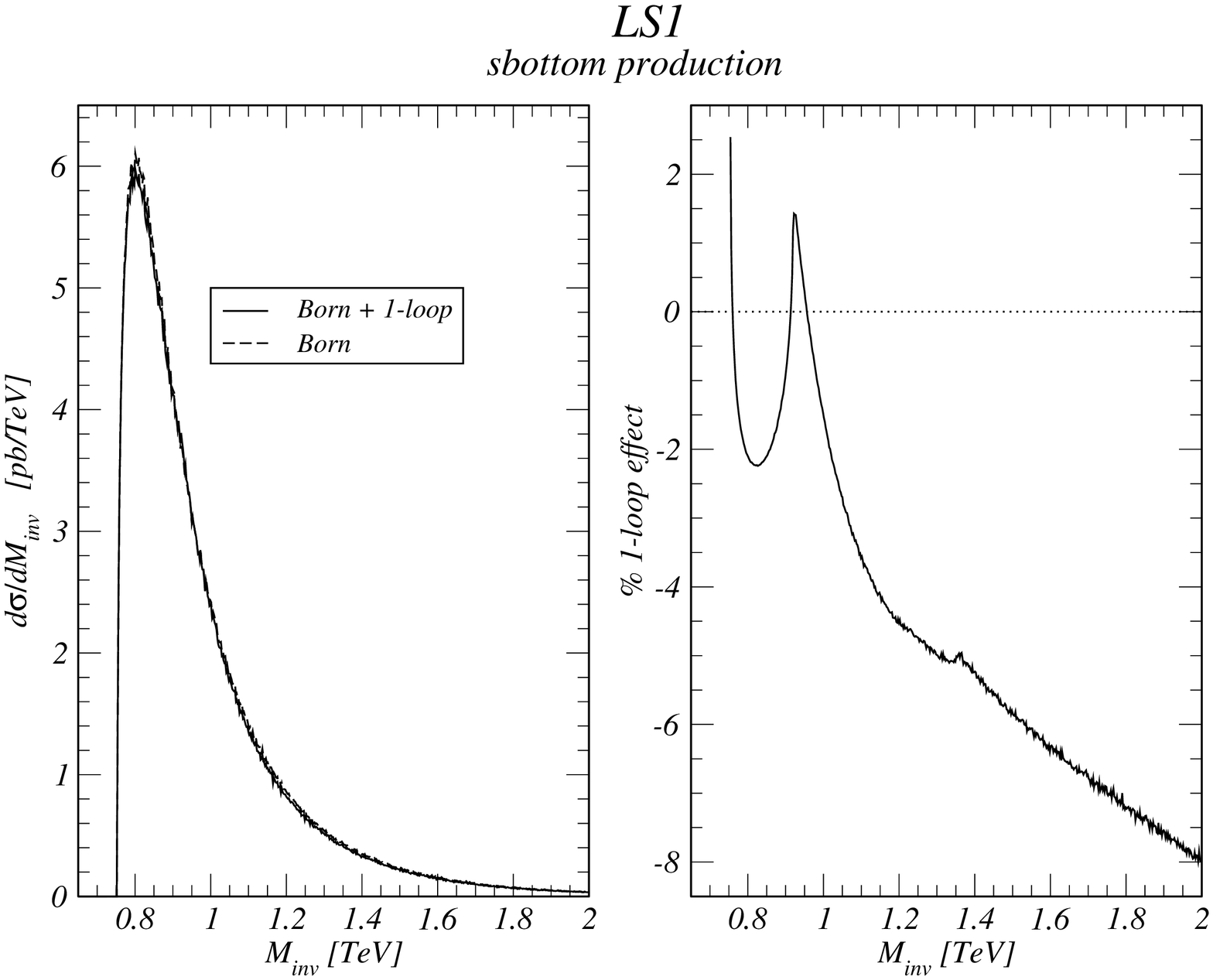, width=18cm, angle=90}
\vspace{1.5cm}
\caption{LS1, Born and one-loop distribution $d\sigma/dM_{\rm inv}$ for sbottom production. The right panel shows the percentual relative effect.}
\label{fig:distrib:LS1sb}
\end{figure}

\begin{figure}
\centering
\epsfig{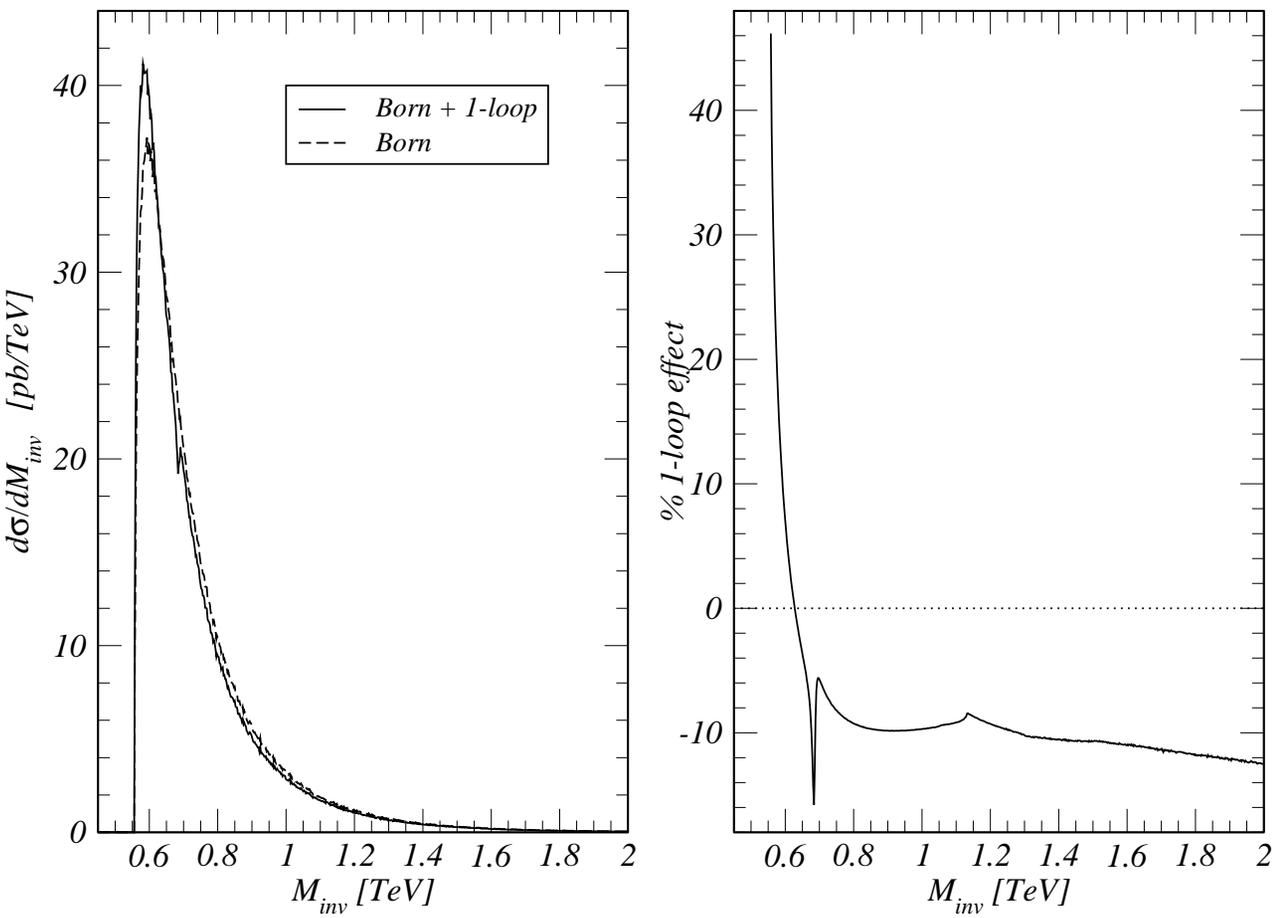}
\vspace{1.5cm}
\caption{SPS5, Born and one-loop distribution $d\sigma/dM_{\rm inv}$ for stop production. The right panel shows the percentual relative effect.}
\label{fig:distrib:SPS5st}
\end{figure}

\begin{figure}
\centering
\epsfig{file=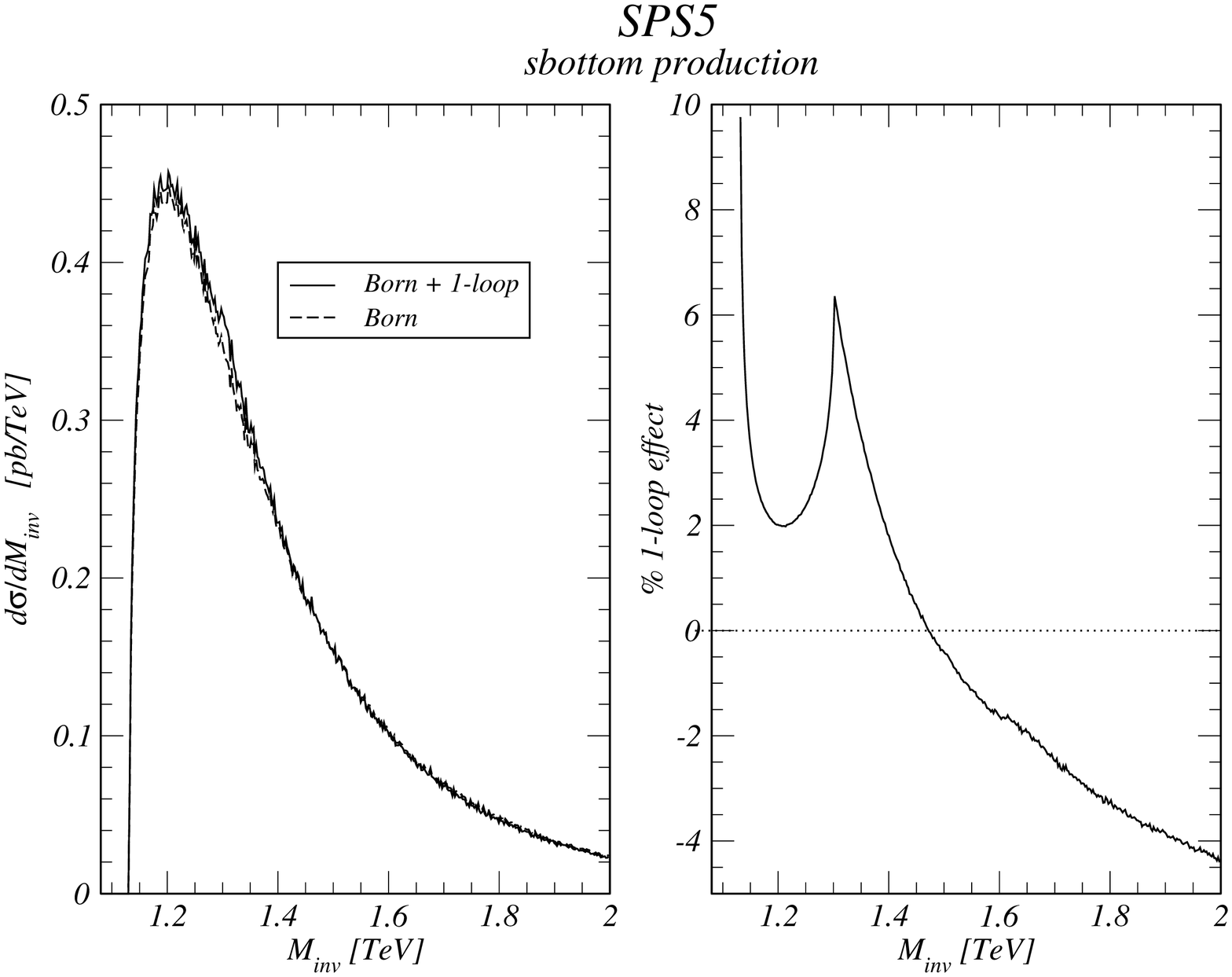, width=18cm, angle=90}
\vspace{1.5cm}
\caption{SPS5, Born and one-loop distribution $d\sigma/dM_{\rm inv}$ for sbottom production. The right panel shows the percentual relative effect.}
\label{fig:distrib:SPS5sb}
\end{figure}

Our results are shown in the next Figures. We have tried to draw a limited number of curves, that contain all the information that seems more relevant to us. With this purpose, we have first shown in Figs.~(\ref{fig:distrib:LS1st},\ref{fig:distrib:LS1sb},\ref{fig:distrib:SPS5st},\ref{fig:distrib:SPS5sb}) the shape of the differential distribution $d\sigma/dM_{\rm inv}$ with the related relative effect for two representative points, chosen as LS1 and SPS5, both for stop and sbottom production. It is possible to see that both for the stop and sbottom cases the relative effect is positive near the threshold, but drops to negative values in the high invariant mass region. The same feature persists in all the remaining considered points.
This can be understood from the discussion of the various helicity amplitudes in Sec.~(II). At large $M_{\rm inv}$, the helicity conserving amplitude dominates with its Sudakov negative correction, while at small $M_{\rm inv}$ the helicity violating amplitude is the larger one and receives a positive correction in a narrow region near the production threshold.

However, due to the different masses of stops and sbottoms and to the different particles involved in the loops, there are substantial differences between the two processes: in the stop case the positive relative effects in the very low mass region soon vanishes, approaching typically a -10\% limit, while in the sbottom case it is possible to note that threshold effects (the peaks and troughs in the low mass region) are more pronounced and produce a typically bigger positive contribution which drops slowly to different limits in the high mass region.
As a consequence, in the stop case one may expect to find a rather small effect in the total rate due to the cancellation between the corrections in these two regimes; in the sbottom case, by contrast, it is not possible, a priori, to predict whether the total one-loop effect will be positive or negative and to what extent, therefore to analyze the corrections to the Born results the numerical evaluation is necessary.

In Tab.~(\ref{tab:diagonal}) we show the numerical values of the total rates for the different benchmark points. To allow a comparison with other calculations, we also show the values of the lighter squark masses that are fixed by the SUSPECT~\cite{Djouadi:2002ze} and FeynHiggs~\cite{Frank:2006yh} codes that we used. \newline

\begin{figure}
\epsfig{file=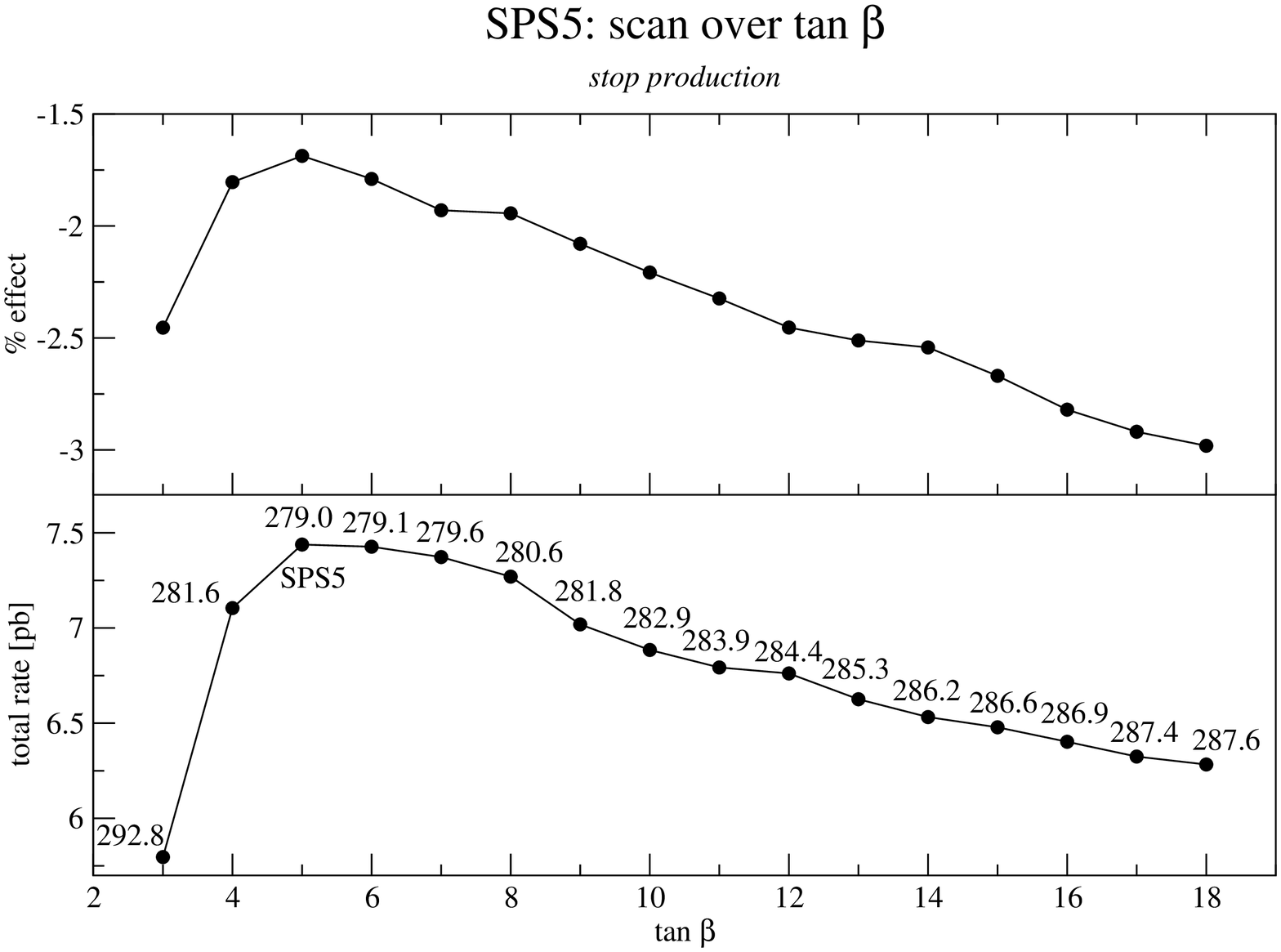, width=0.88\textwidth, angle=0}
\epsfig{file=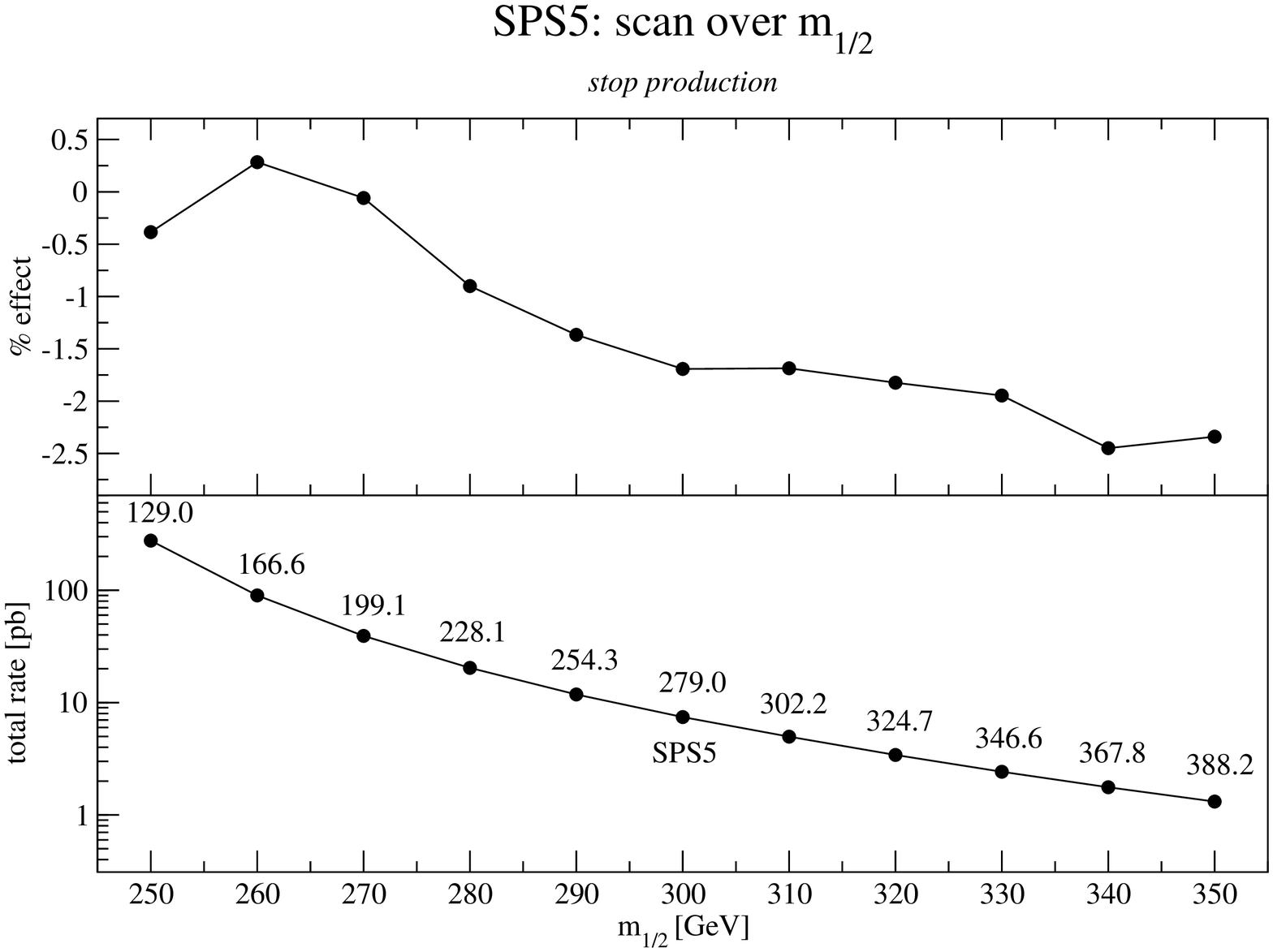, width=0.88\textwidth, angle=0}
\caption{SPS5: scans over the mSUGRA parameters $\tan\beta$ and $m_{1/2}$ for diagonal stop production. The top panels show the percentual effect on the integrated cross section, the bottom panels show the variation in the value of the total cross section; the numbers above the curves in the bottom panels represent the value of the stop mass $m_{\tilde t_1}$(in GeV).}
\label{fig:tanbetam12SPS5}
\end{figure}

\begin{figure}
\epsfig{file=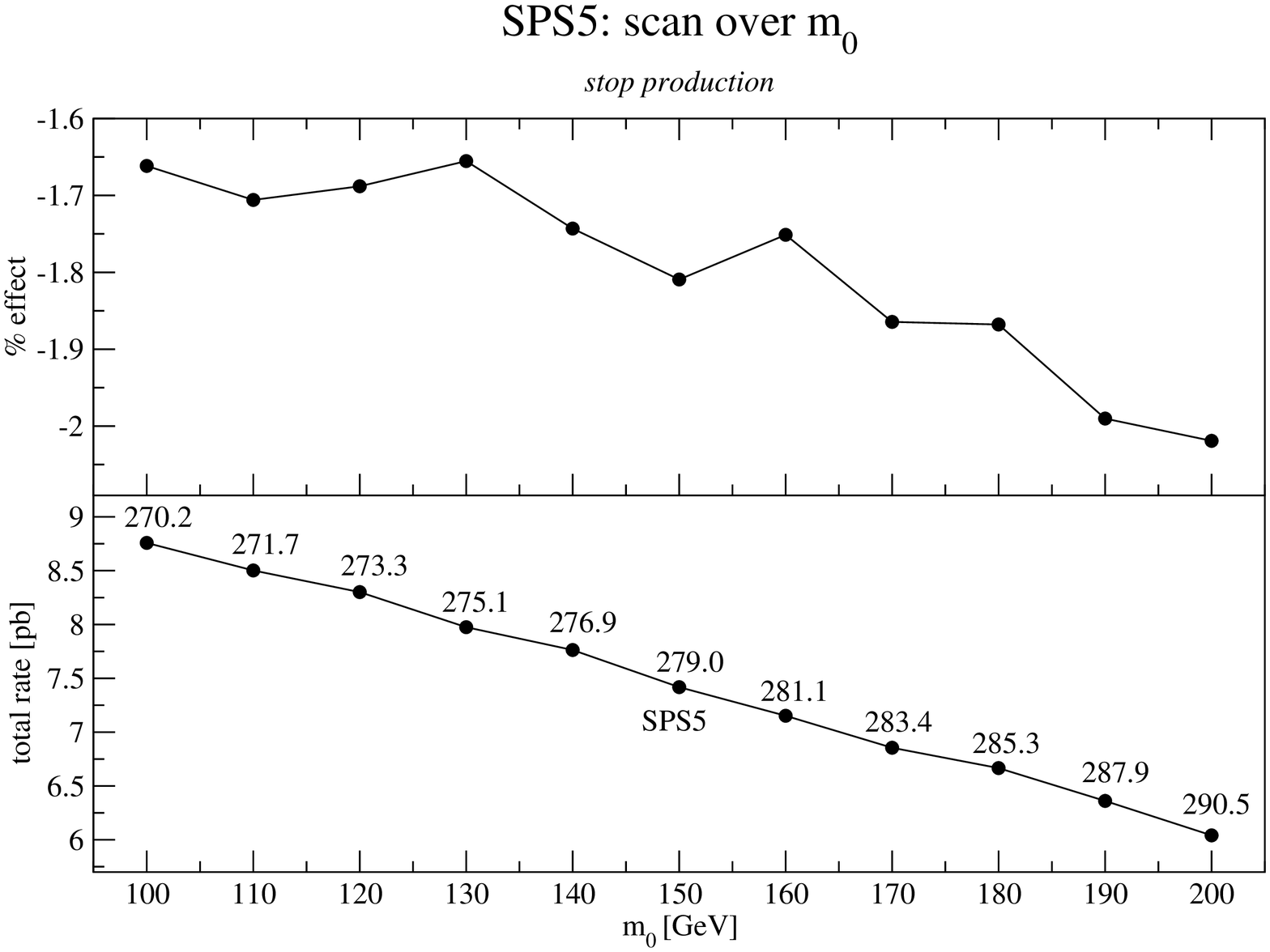, width=0.88\textwidth, angle=0}
\epsfig{file=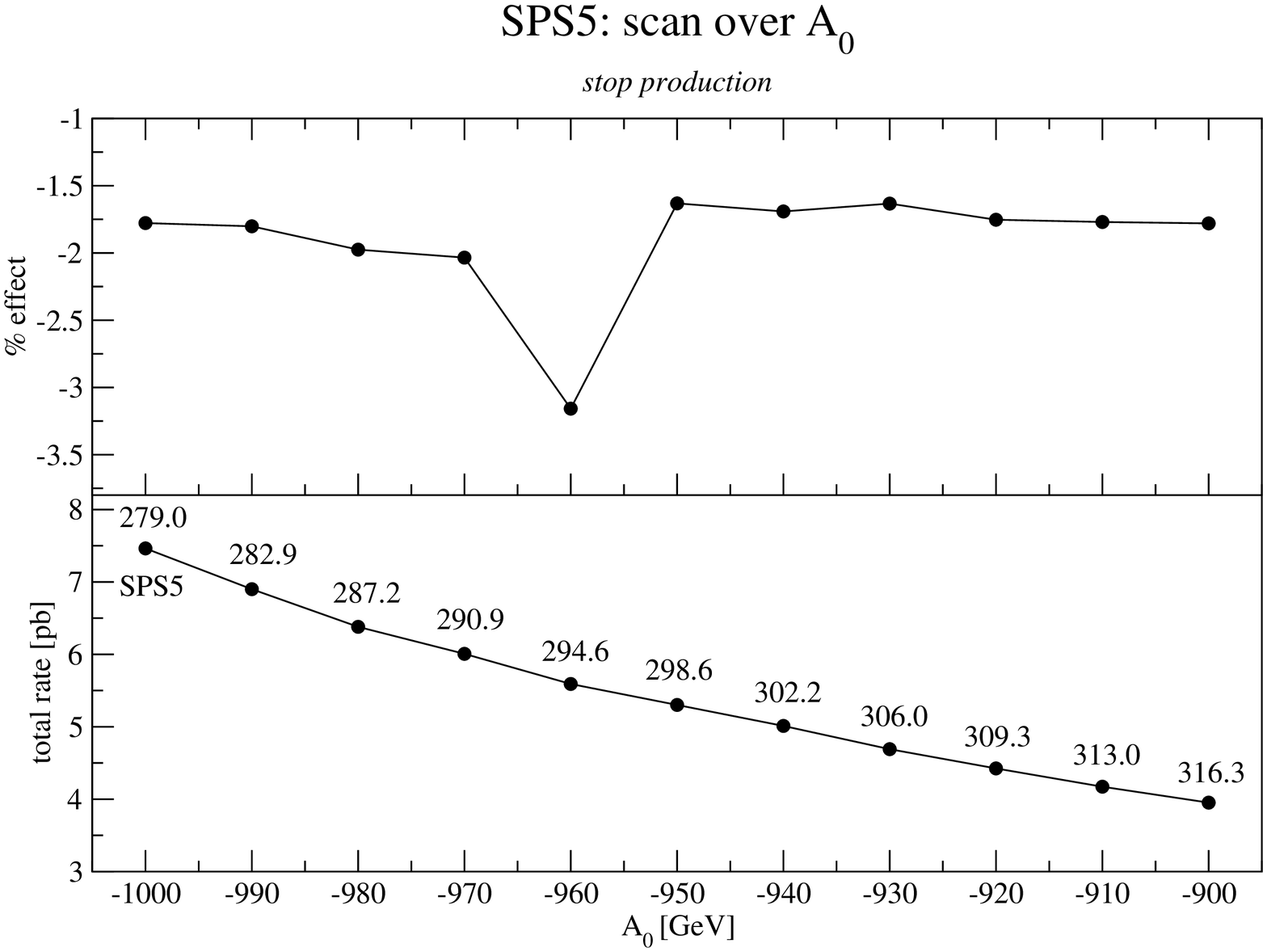, width=0.88\textwidth, angle=0}
\caption{SPS5: scans over the mSUGRA parameters $m_0$ and $A_0$ for diagonal stop production. The top panels show the percentual effect on the integrated cross section, the bottom panels show the variation in the value of the total cross section; the numbers above the curves in the bottom panels represent the value of the stop mass $m_{\tilde t_1}$(in GeV).}
\label{fig:m0A0SPS5}
\end{figure}

\begin{figure}
\epsfig{file=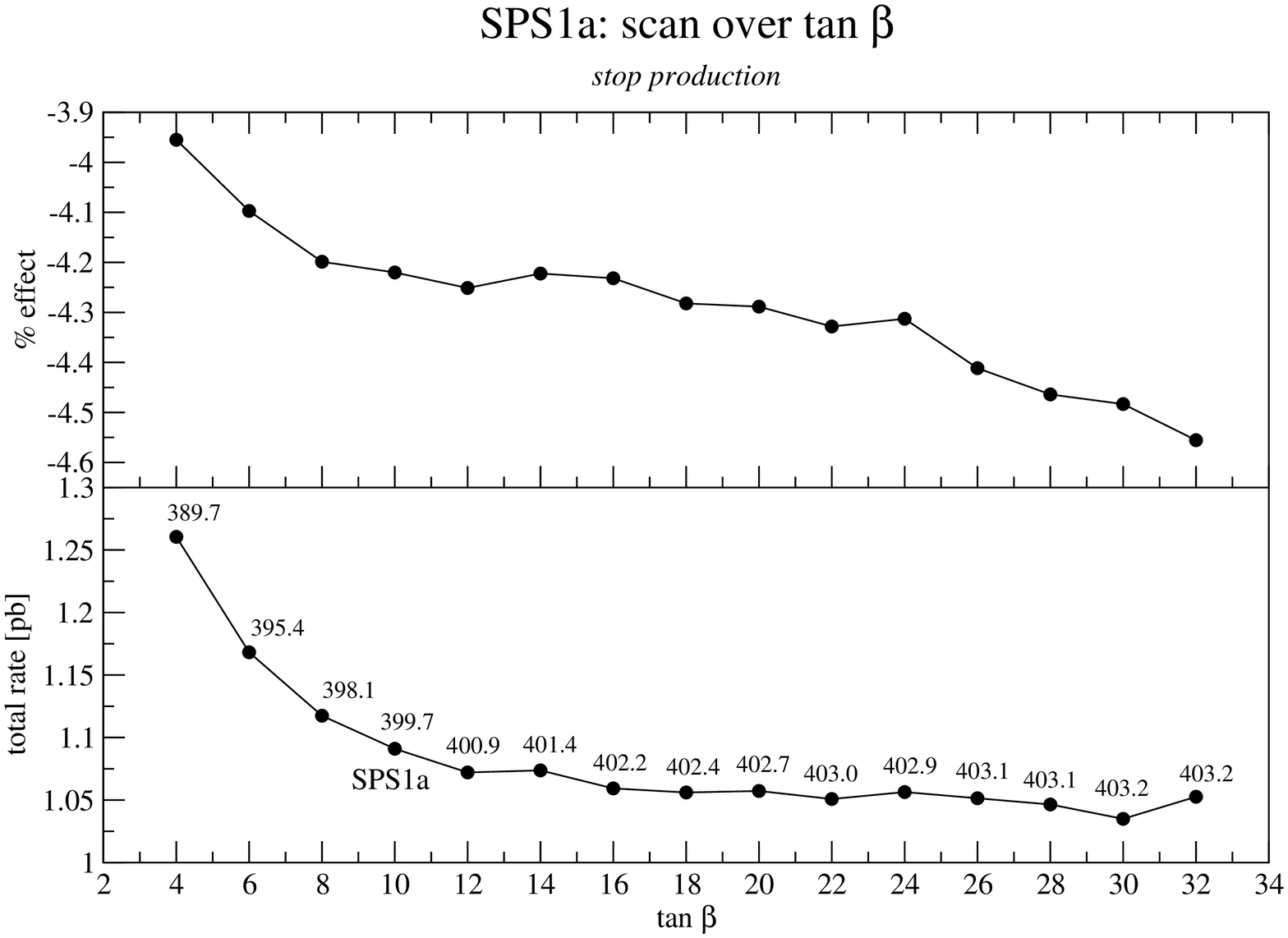, width=0.88\textwidth, angle=0}
\epsfig{file=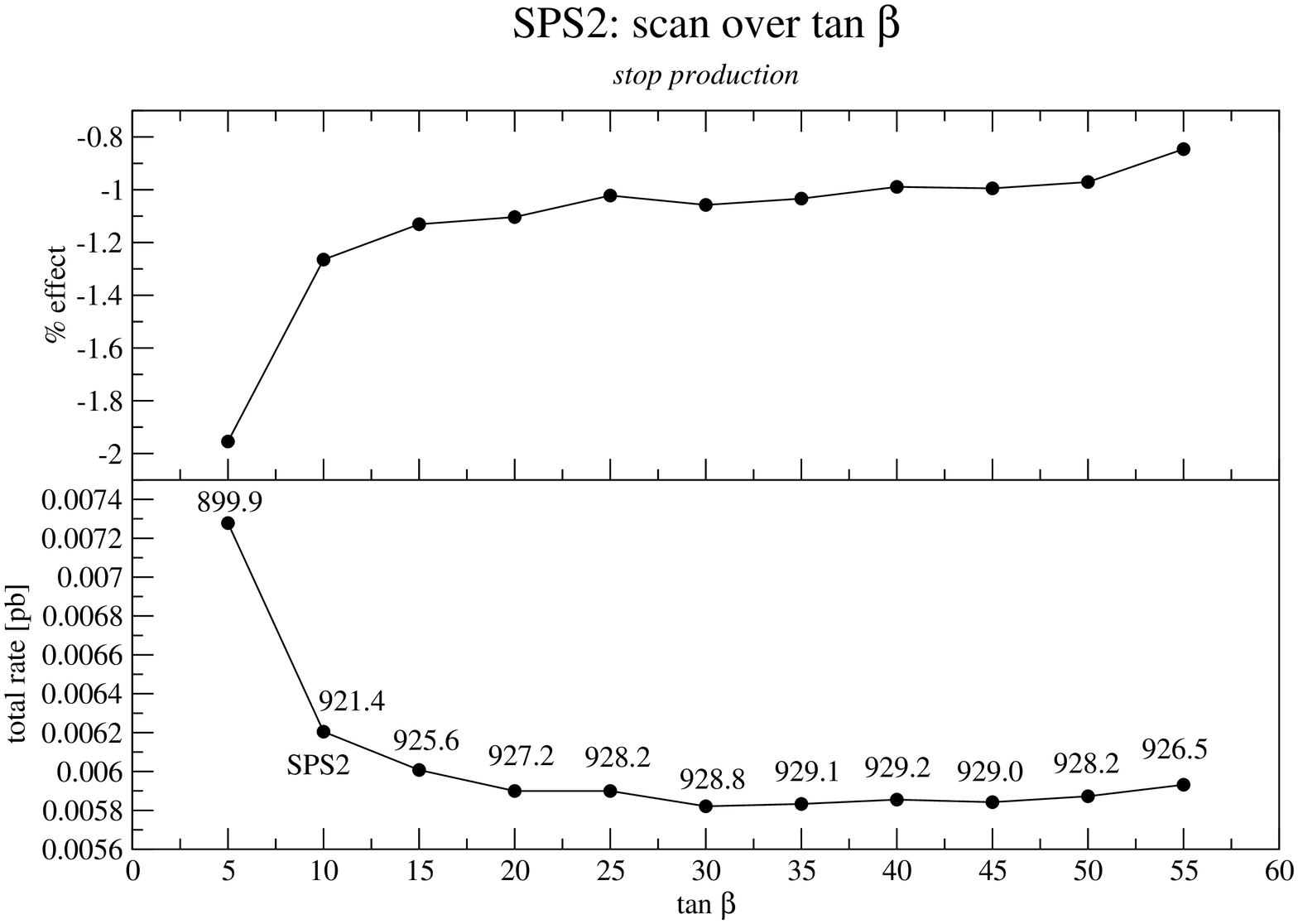, width=0.88\textwidth, angle=0}
\caption{SPS1a and SPS2: scan over the mSUGRA parameter $\tan\beta$ for diagonal stop production. The top panels show the percentual effect on the integrated cross section, the bottom panels show the variation in the value of the total cross section; the numbers above the curves in the bottom panels represent the value of the stop mass $m_{\tilde t_1}$(in GeV).}
\label{fig:tanbetaSPS1SPS2}
\end{figure}

\begin{figure}
\epsfig{file=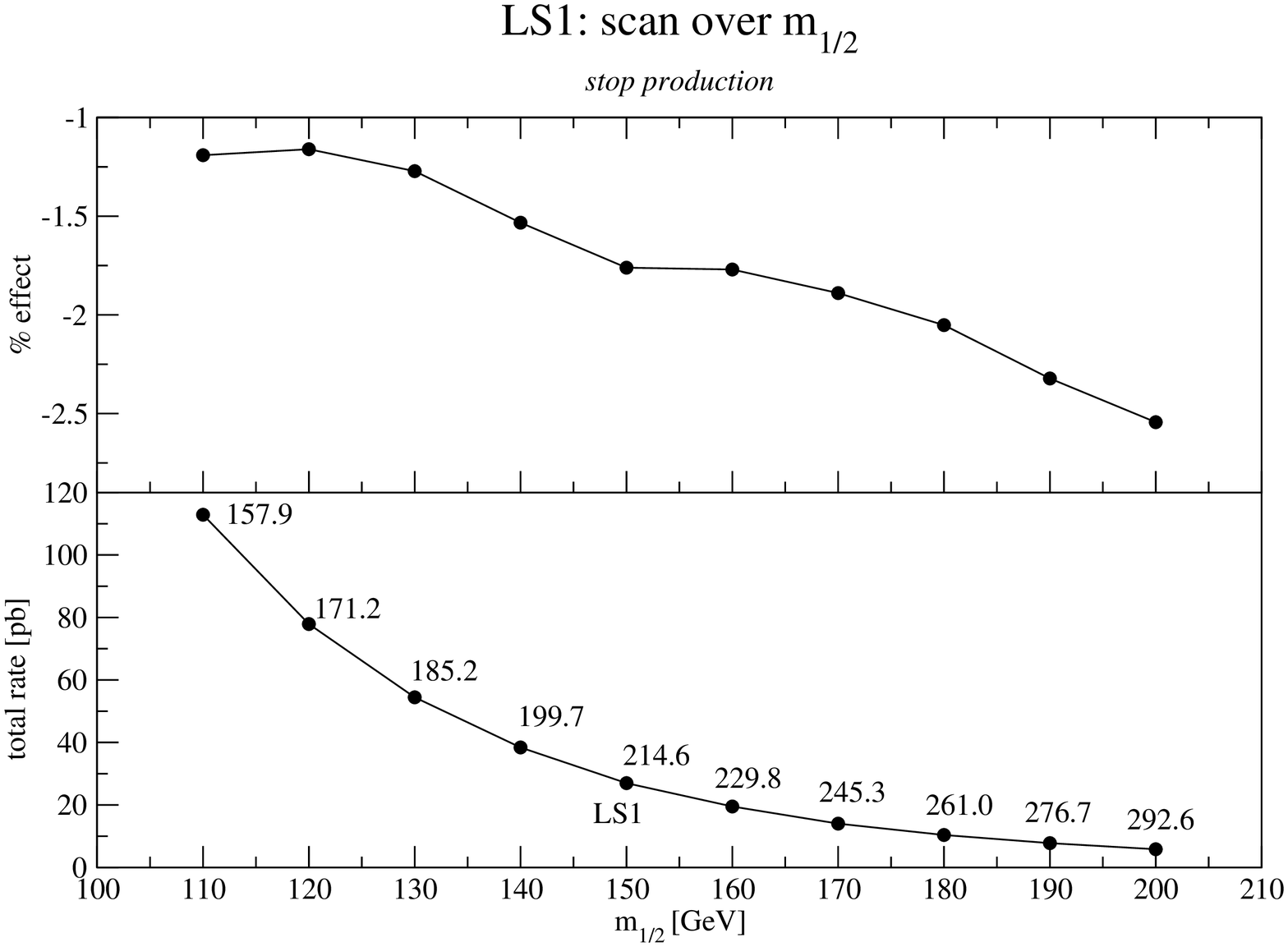, width=0.88\textwidth, angle=0}
\epsfig{file=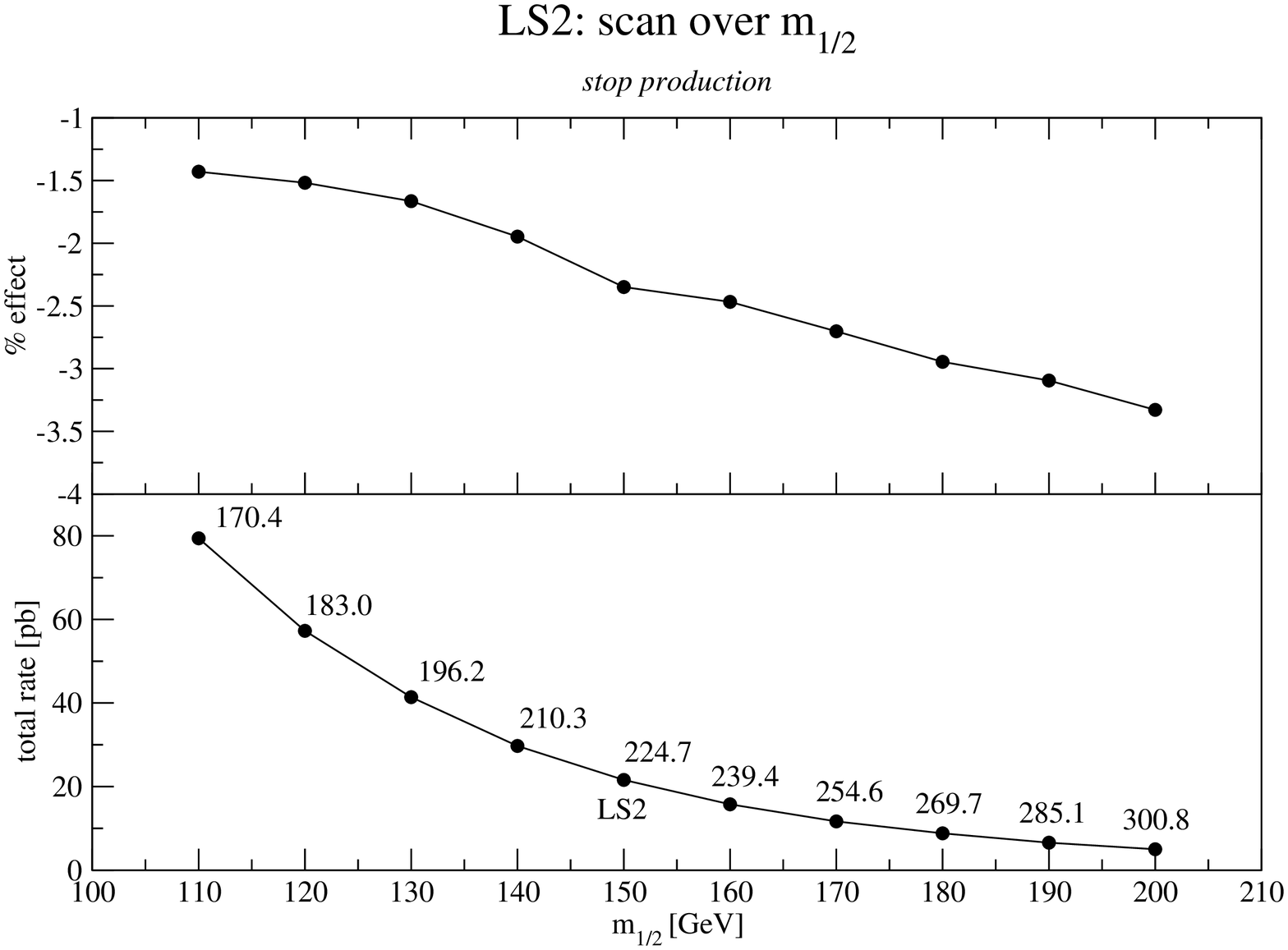, width=0.88\textwidth, angle=0}
\caption{LS1 and LS2: scan over the mSUGRA parameter $m_{1/2}$ for diagonal stop production. The top panels show the percentual effect on the integrated cross section, the bottom panels show the variation in the value of the total cross section; the numbers above the curves in the bottom panels represent the value of the stop mass $m_{\tilde t_1}$(in GeV).}
\label{fig:m12LS1LS2}
\end{figure}

\begin{figure}
\epsfig{file=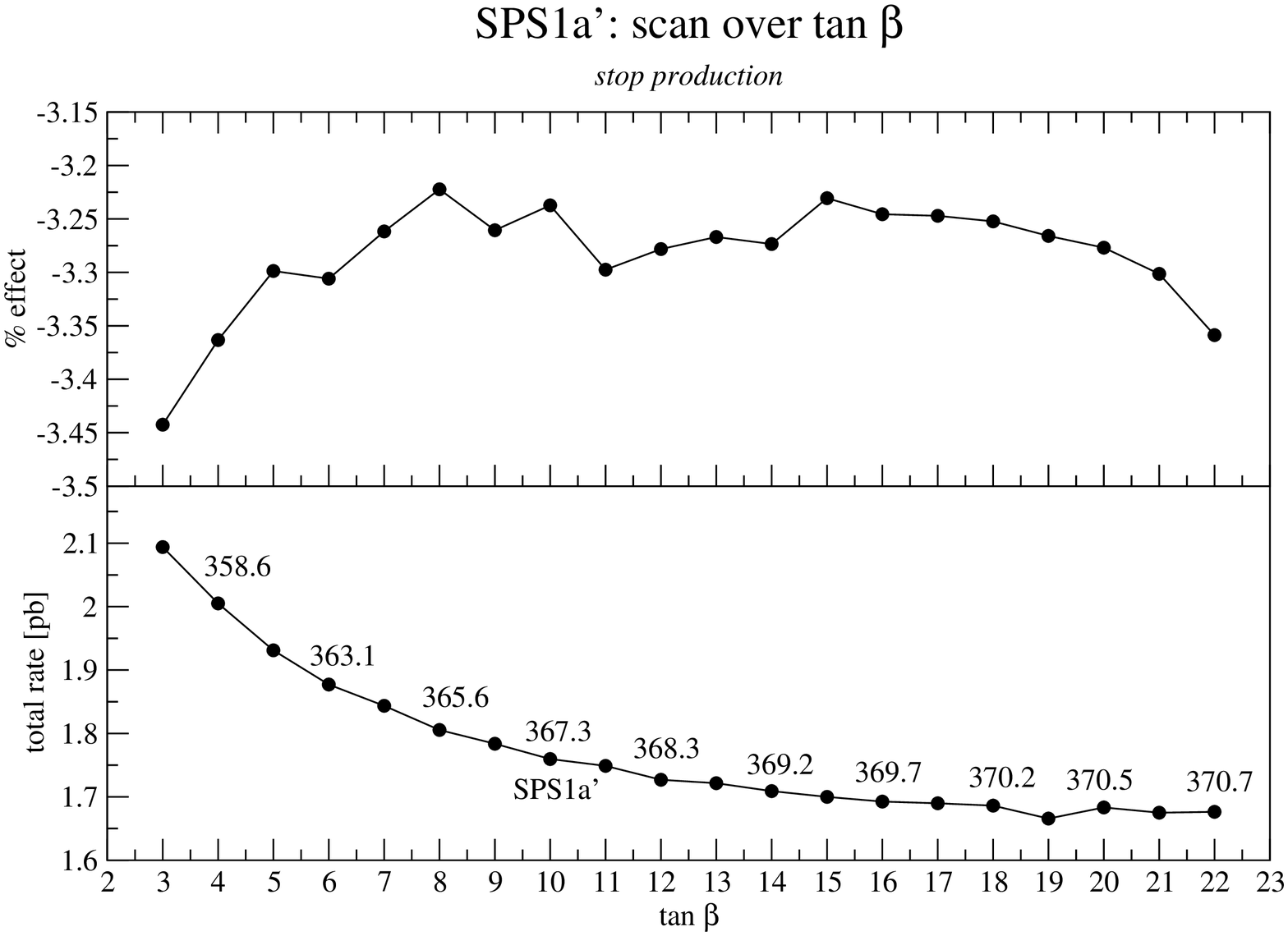, width=0.88\textwidth, angle=0}
\epsfig{file=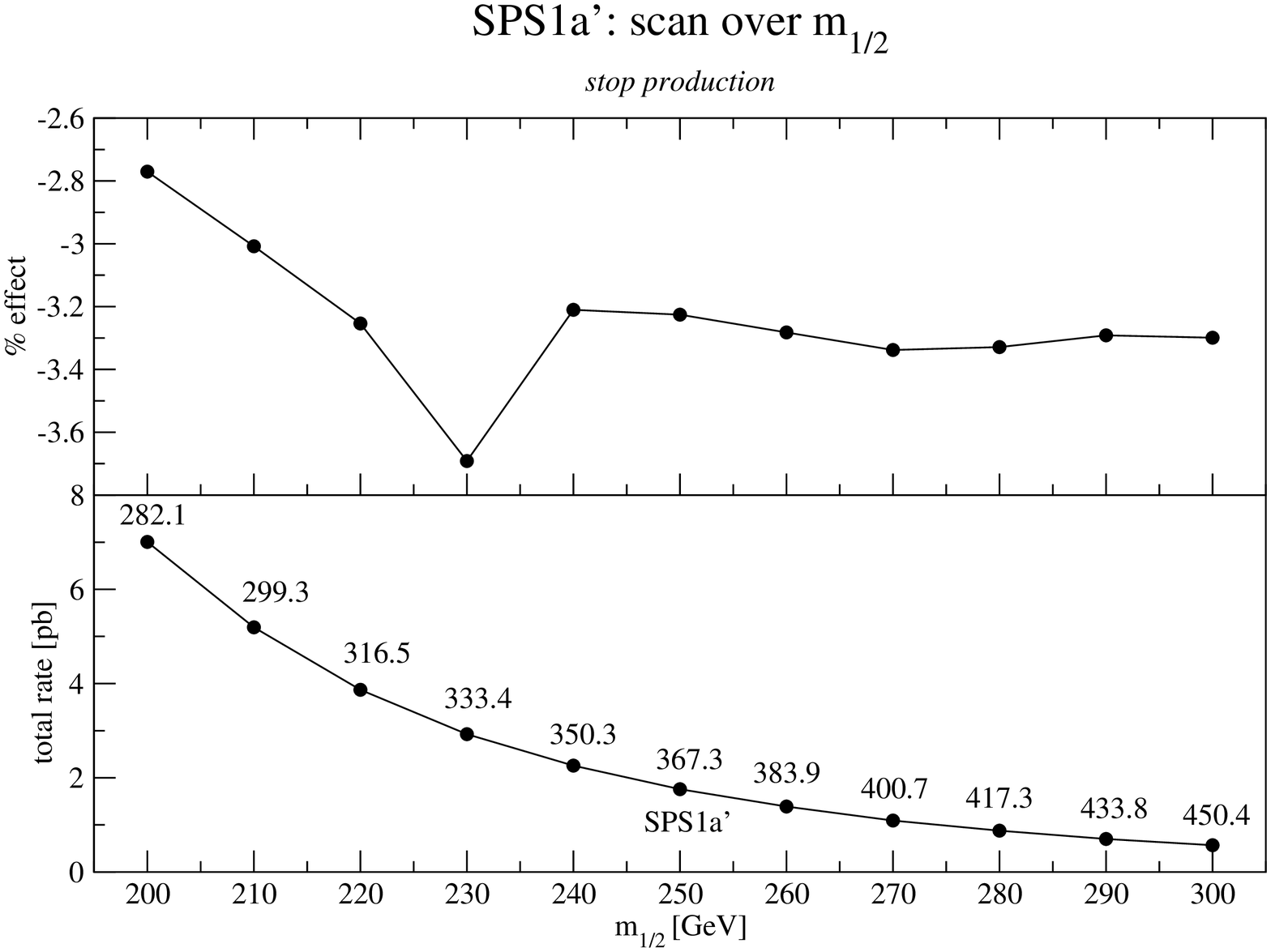, width=0.88\textwidth, angle=0}
\caption{SPS1a': scan over the mSUGRA parameters $\tan\beta$ and $m_{1/2}$ for diagonal stop production. The top panels show the percentual effect on the integrated cross section, the bottom panels show the variation in the value of the total cross section; the numbers above the curves in the bottom panels represent the value of the stop mass $m_{\tilde t_1}$(in GeV).}
\label{fig:tanbetam12SPS1aprime}
\end{figure}

\begin{figure}
\epsfig{file=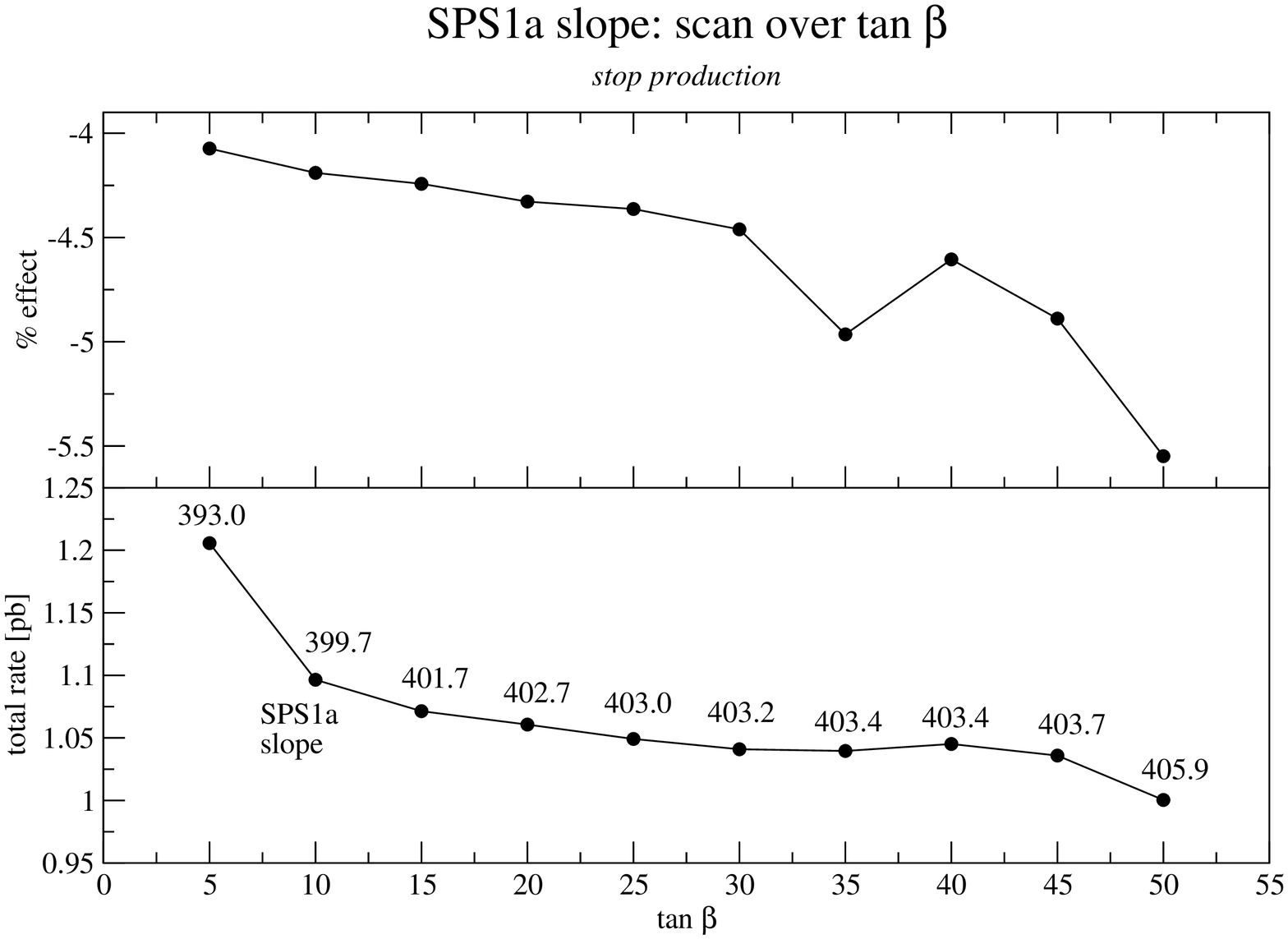, width=0.88\textwidth, angle=0}
\epsfig{file=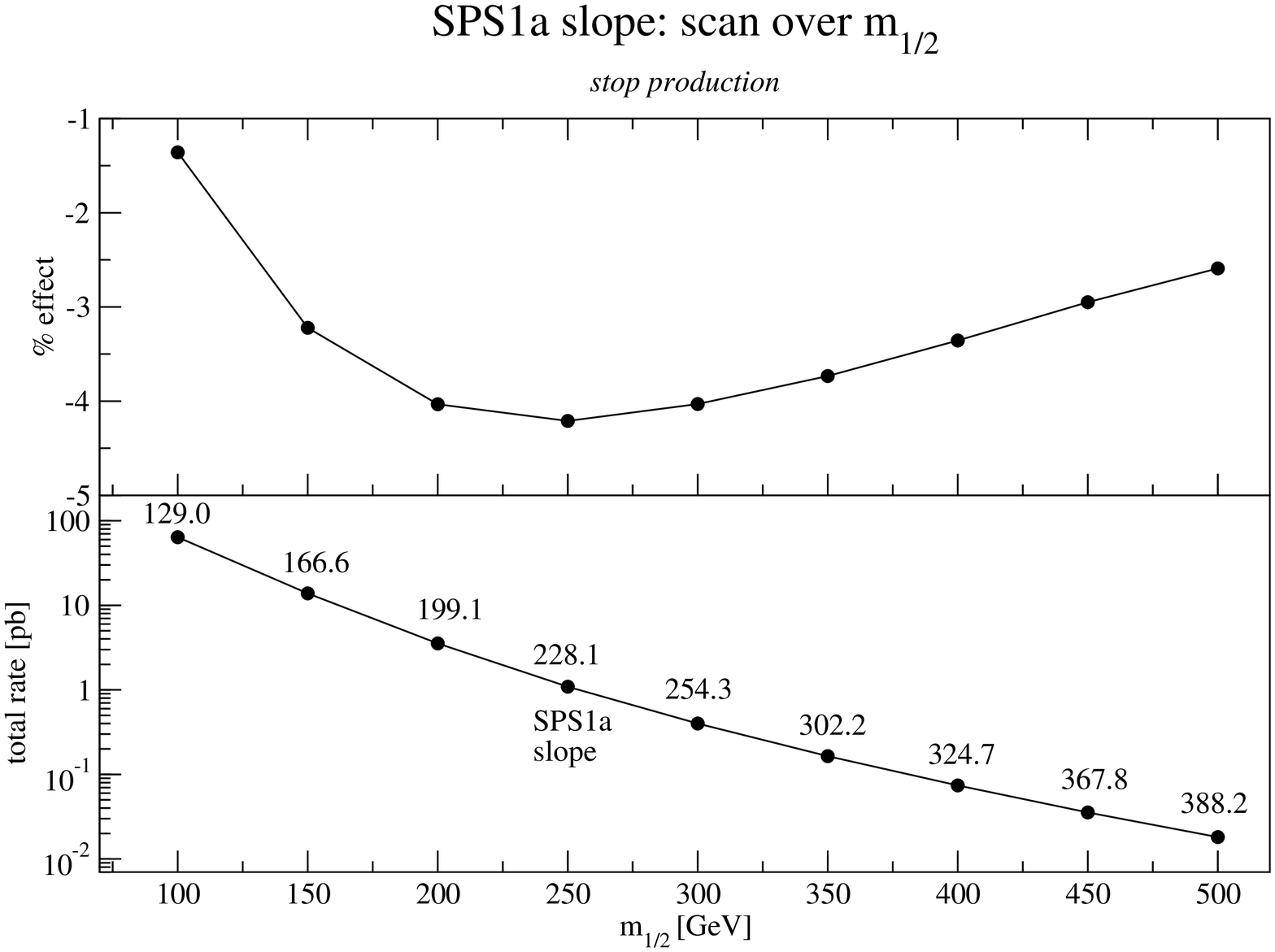, width=0.88\textwidth, angle=0}
\caption{SPS1a slope: scan over the mSUGRA parameter $\tan\beta$ and $m_{1/2}$ for diagonal stop production. The top panels show the percentual effect on the integrated cross section, the bottom panels show the variation in the value of the total cross section; the numbers above the curves in the bottom panels represent the value of the stop mass $m_{\tilde t_1}$(in GeV).}
\label{fig:tanbetam12SPS1aslope}
\end{figure}

\begin{figure}
  \subfigure{
  \begin{minipage}{0.5\textwidth}
    \epsfig{file=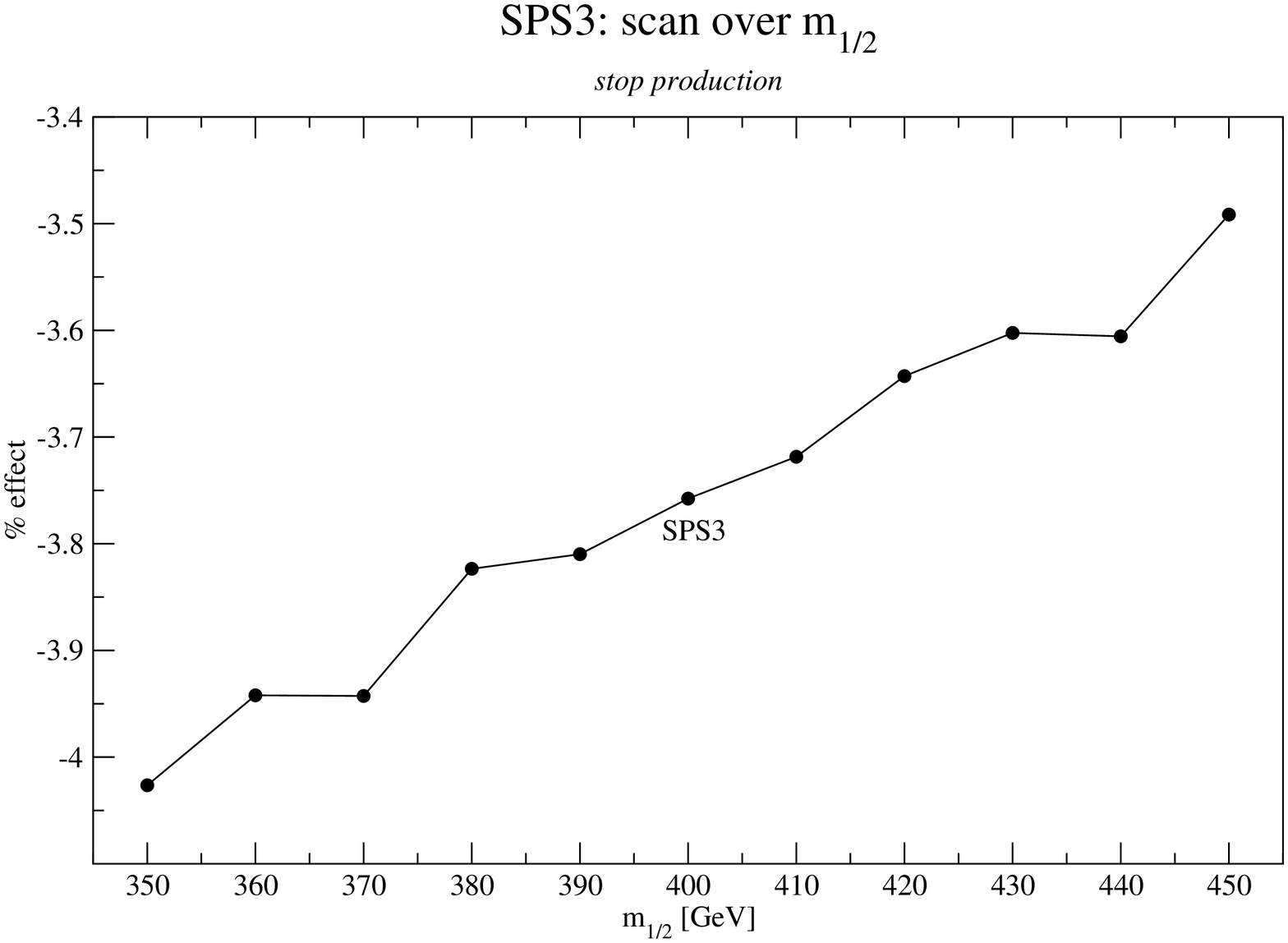, width=\textwidth, angle=0}
    \epsfig{file=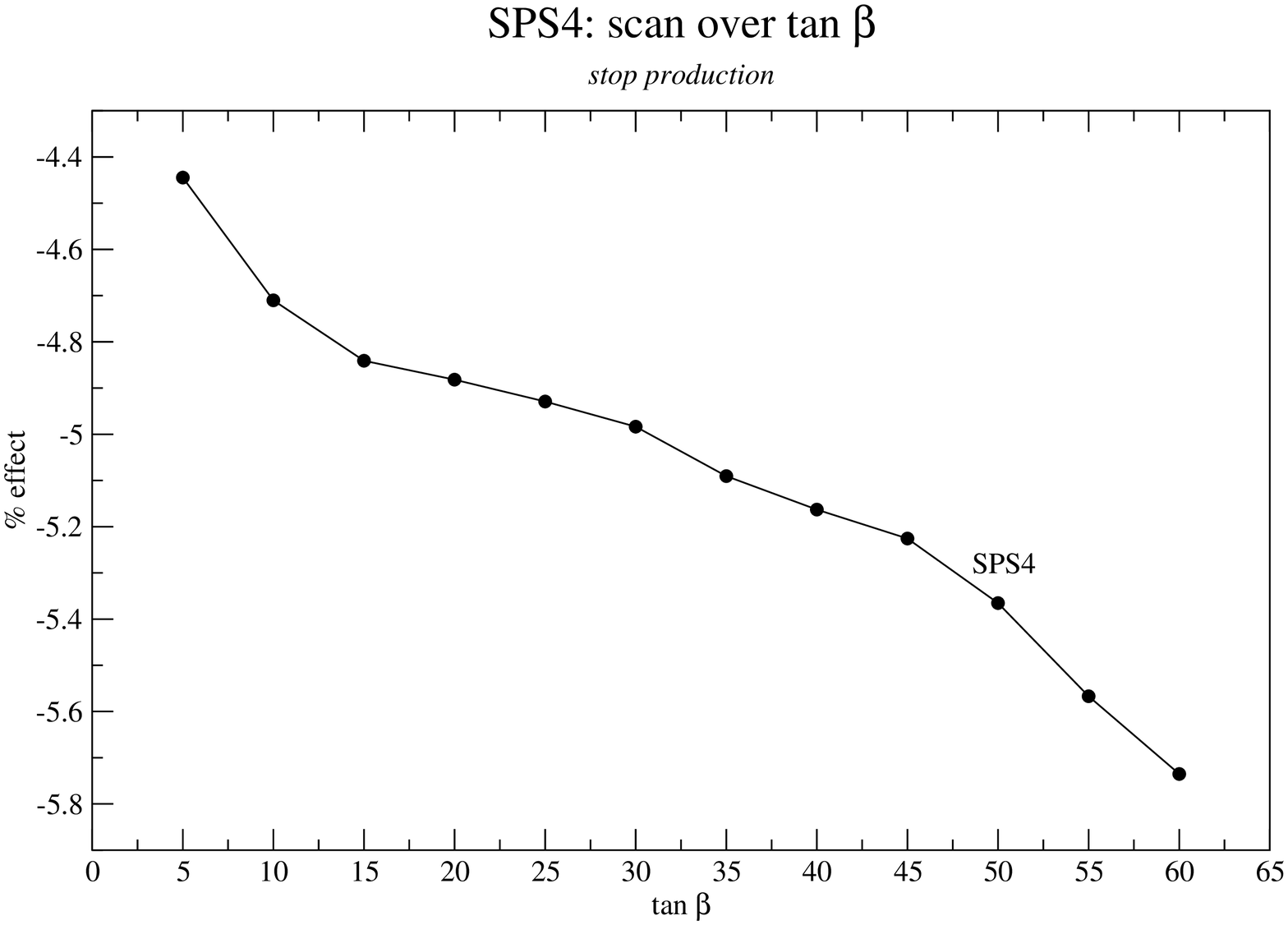, width=\textwidth, angle=0}
  \end{minipage}
  \begin{minipage}{0.5\textwidth}
    \epsfig{file=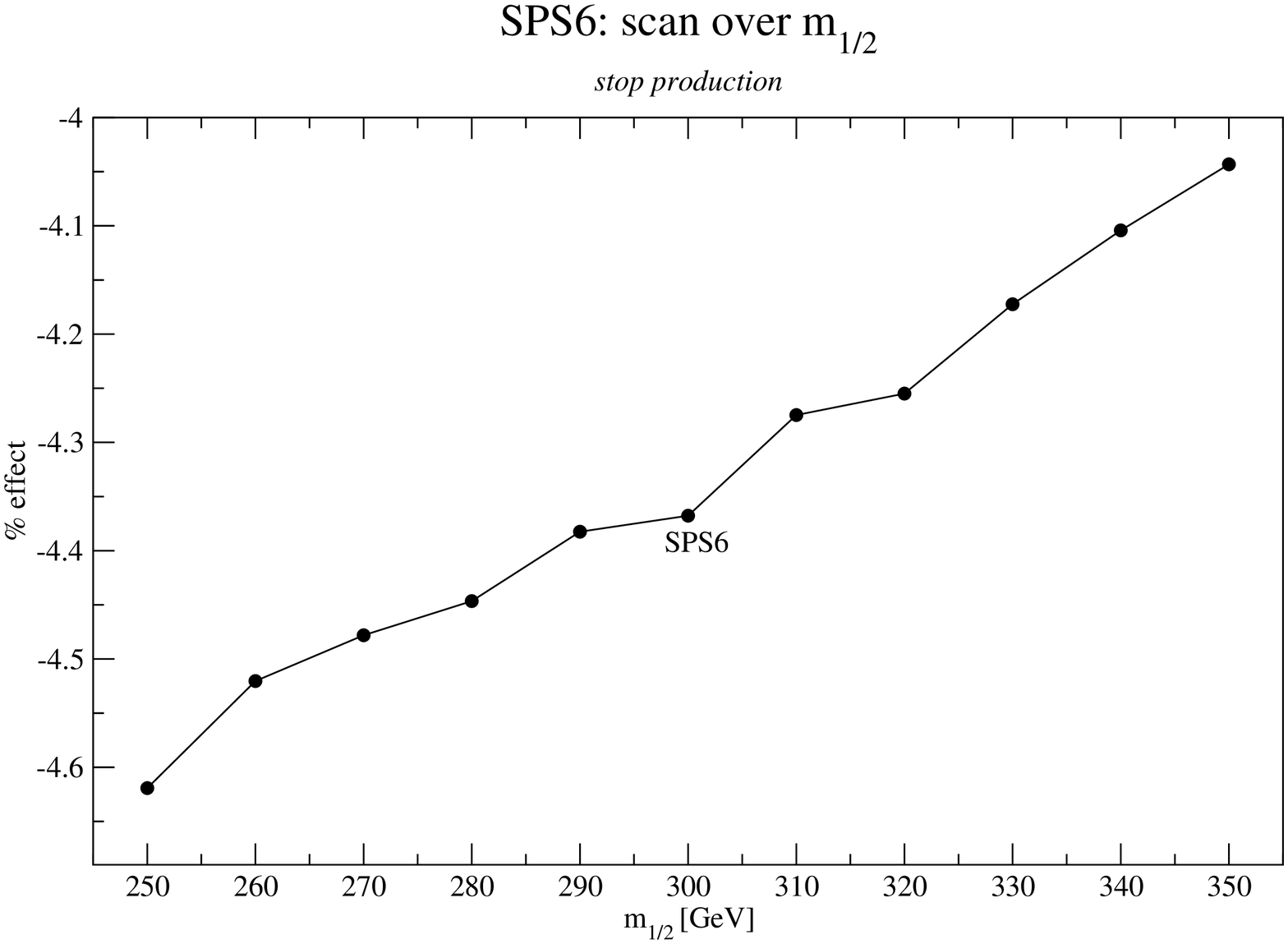, width=\textwidth, angle=0}
    \epsfig{file=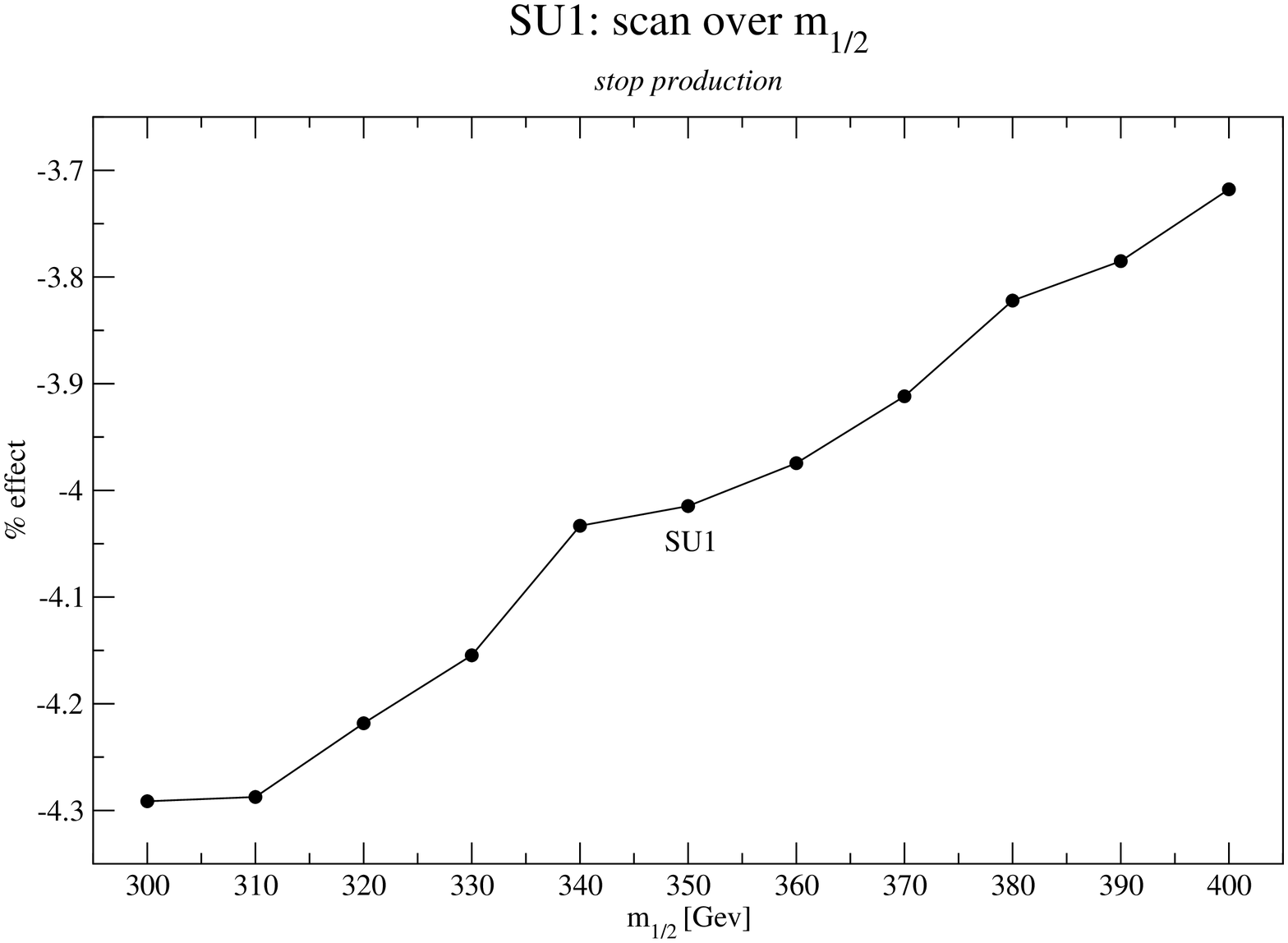, width=\textwidth, angle=0}
  \end{minipage}}
  \epsfig{file=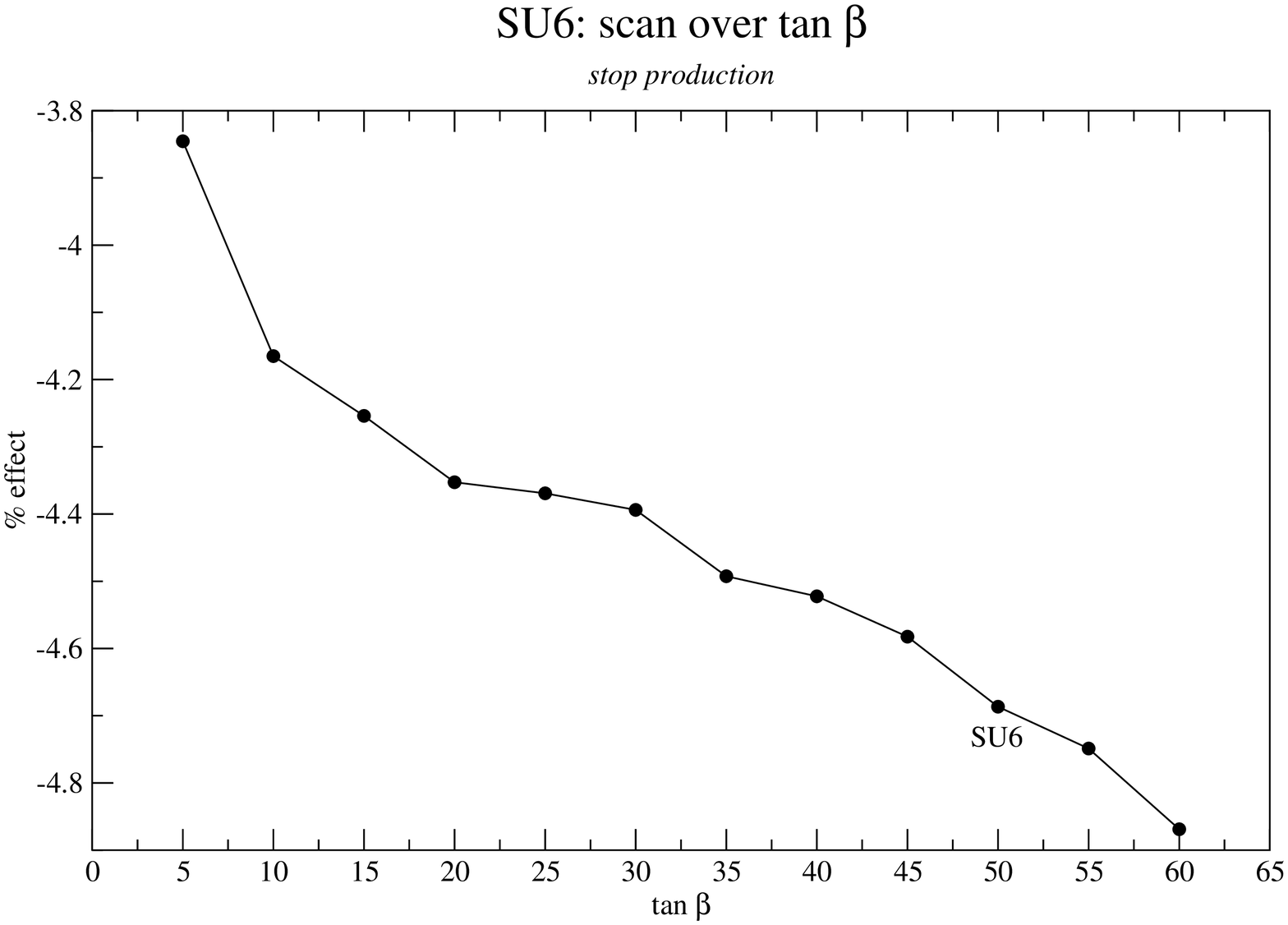, width=0.5\textwidth, angle=0}
\caption{Dominant parameter dependence on one loop effects for the benchmark points with small cross section for diagonal stop production.}
\label{fig:remainingbenchmarks}
\end{figure}

\begin{figure}
\epsfig{file=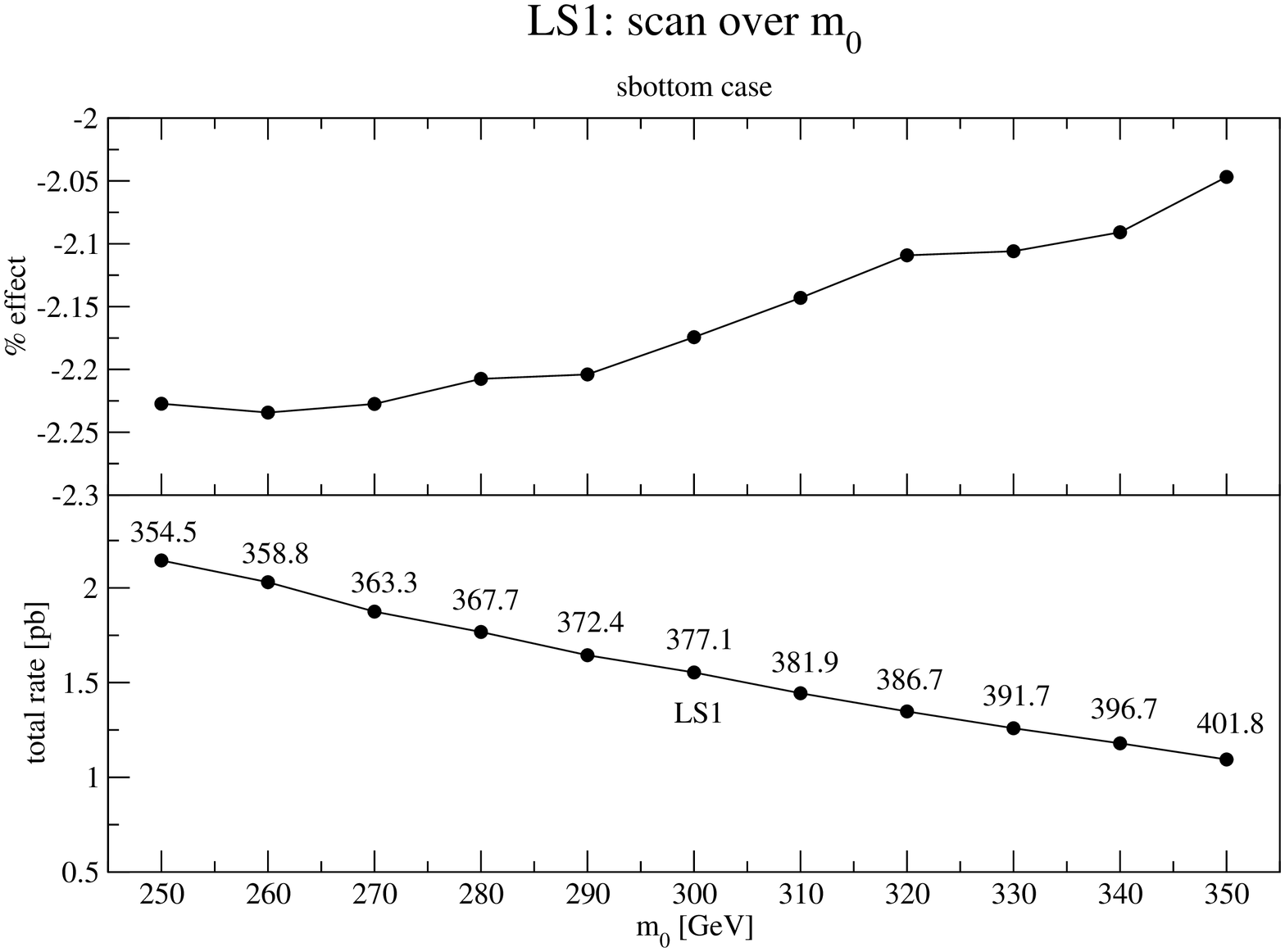, width=0.88\textwidth, angle=0}
\epsfig{file=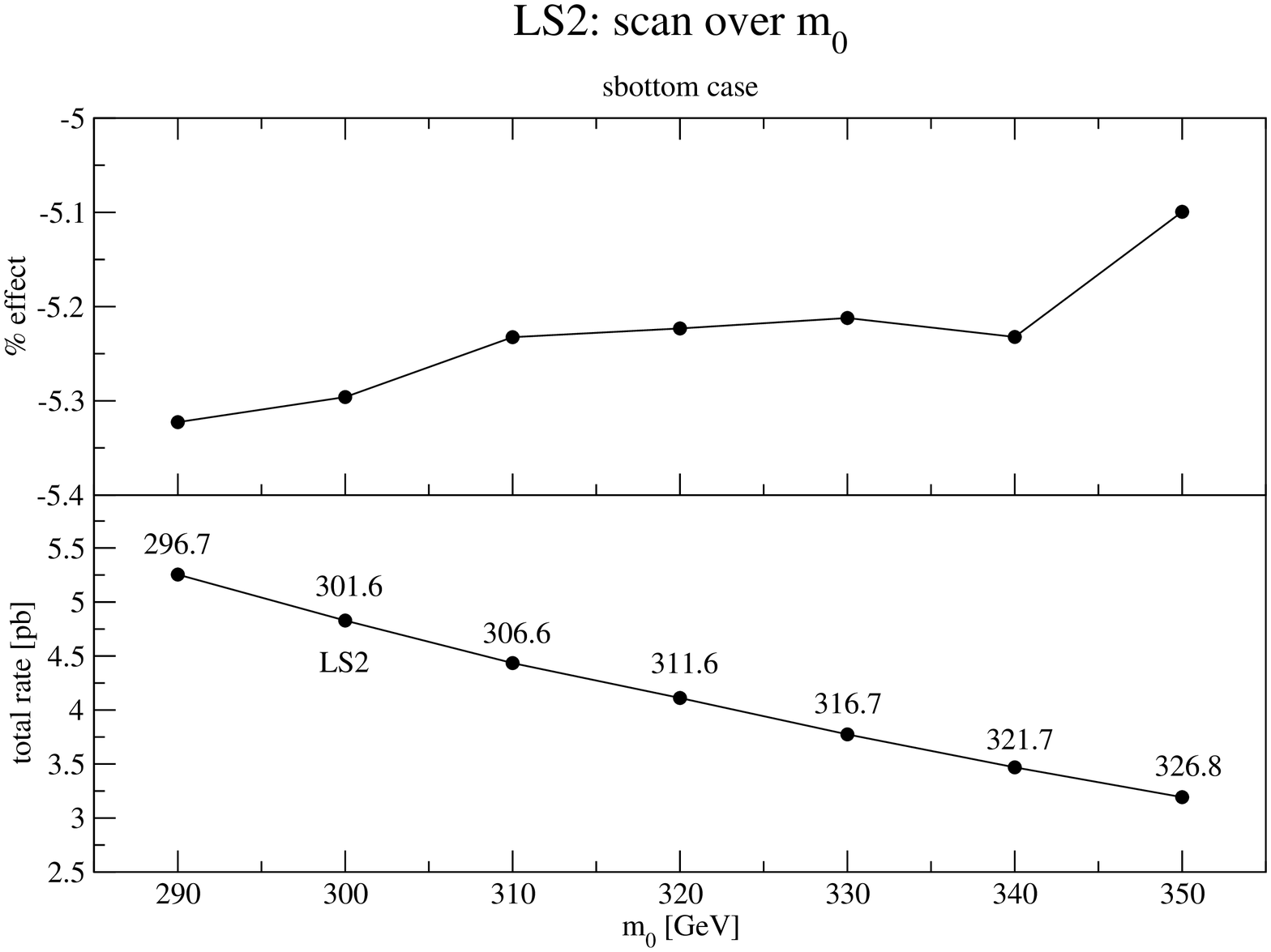, width=0.88\textwidth, angle=0}
\caption{LS1 and LS2: scan over the mSUGRA parameter $m_0$ for diagonal sbottom production. The top panels show the percentual effect on the integrated cross section, the bottom panels show the variation in the value of the total cross section; the numbers above the curves in the bottom panels represent the value of the sbottom mass $m_{\tilde b_1}$(in GeV).}
\label{fig:m0sbottom}
\end{figure}

\begin{figure}
\epsfig{file=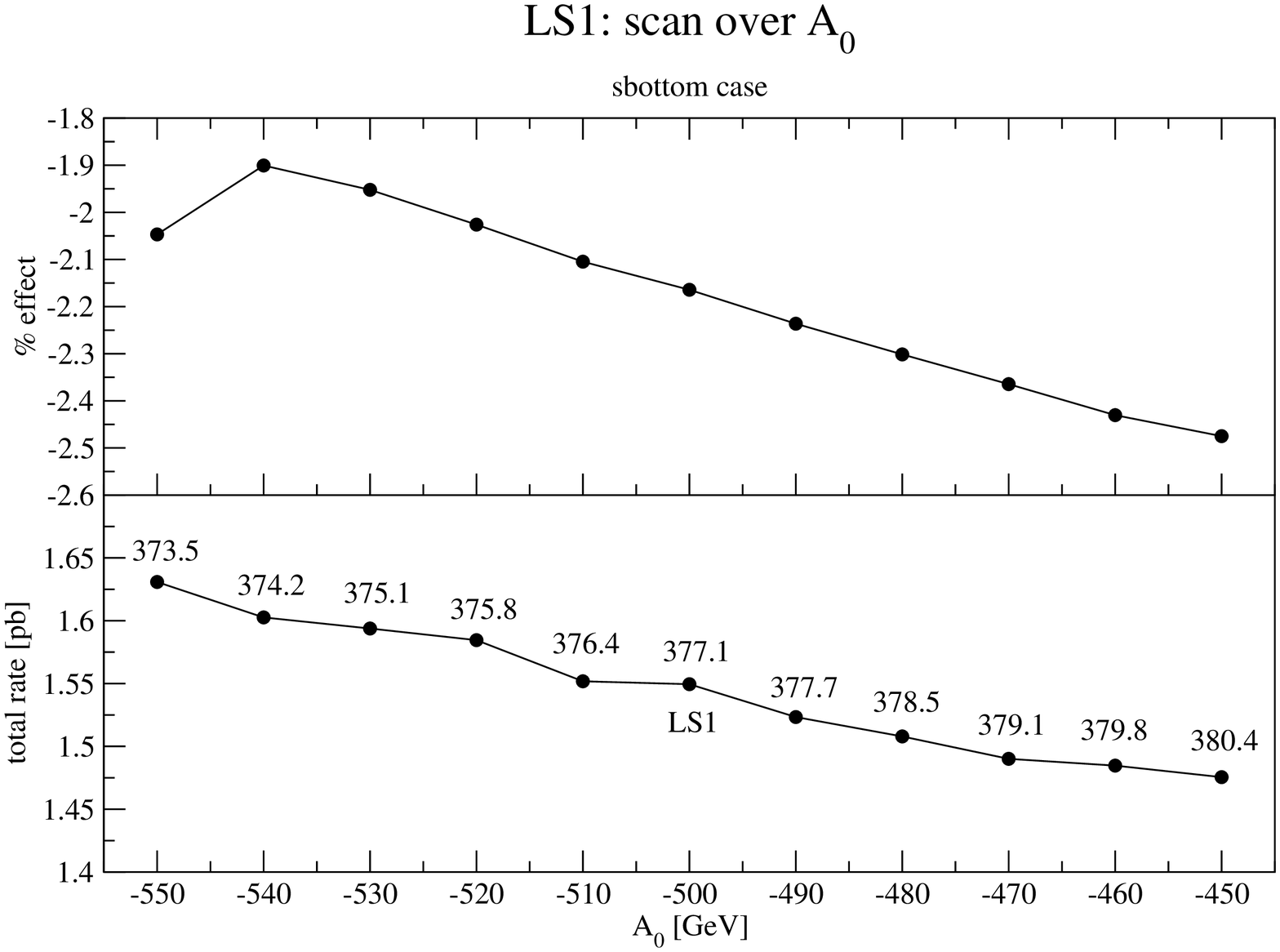, width=0.88\textwidth, angle=0}
\epsfig{file=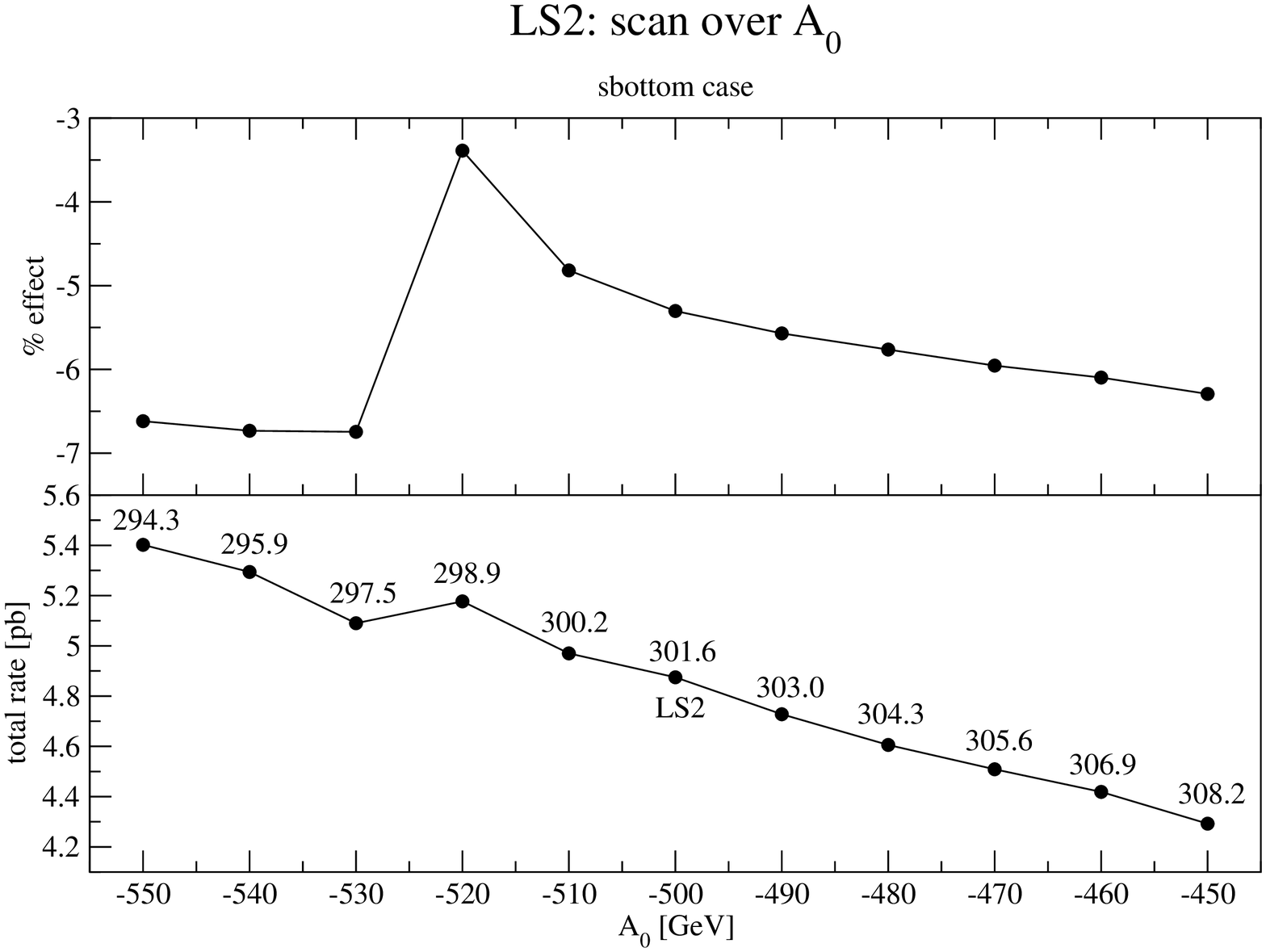, width=0.88\textwidth, angle=0}
\caption{LS1 and LS2: scan over the mSUGRA parameter $A_0$ for diagonal sbottom production. The top panels show the percentual effect on the integrated cross section, the bottom panels show the variation in the value of the total cross section; the numbers above the curves in the bottom panels represent the value of the sbottom mass $m_{\tilde b_1}$(in GeV).}
\label{fig:A0sbottom}
\end{figure}

\begin{figure}
\epsfig{file=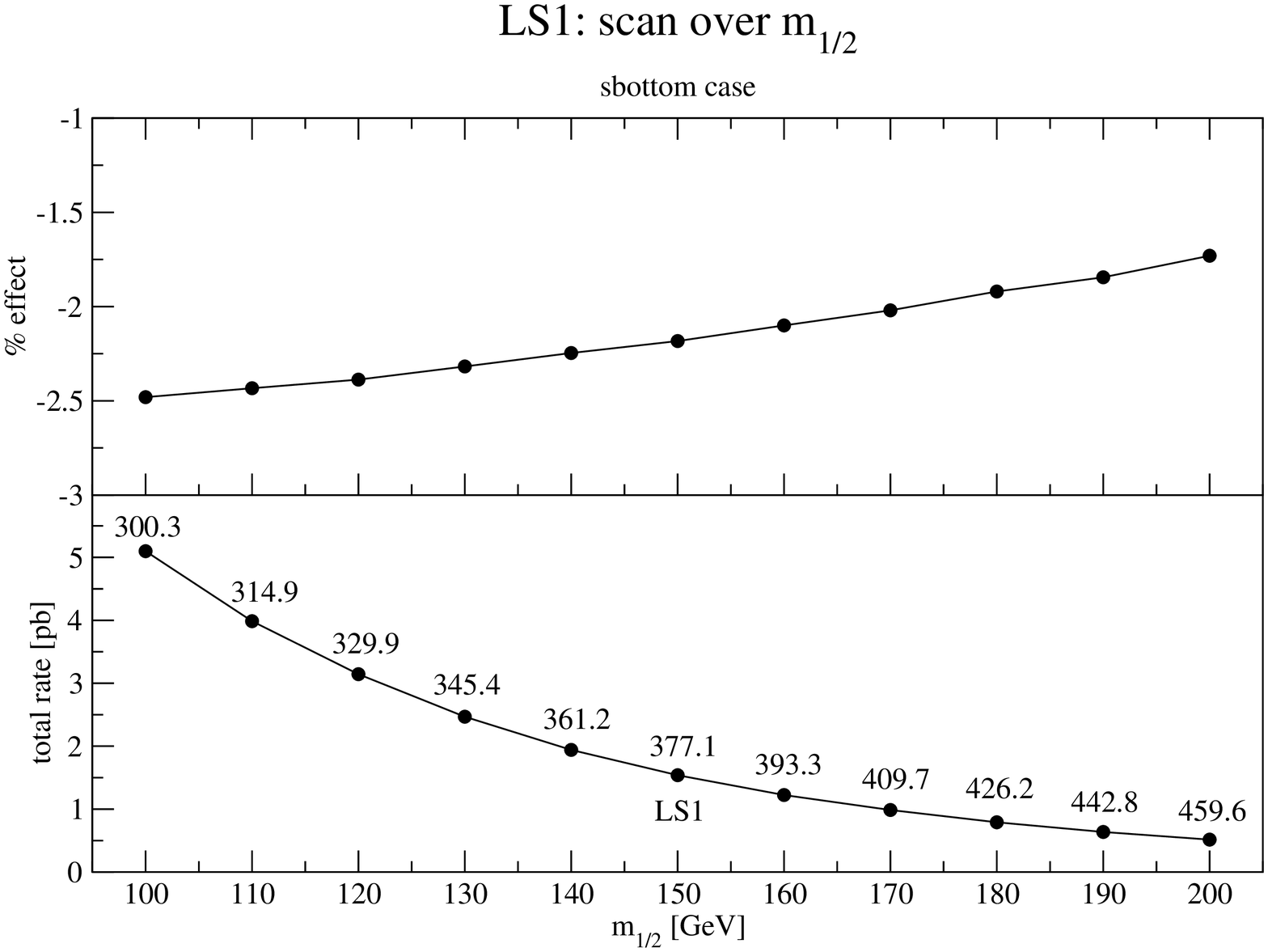, width=0.88\textwidth, angle=0}
\epsfig{file=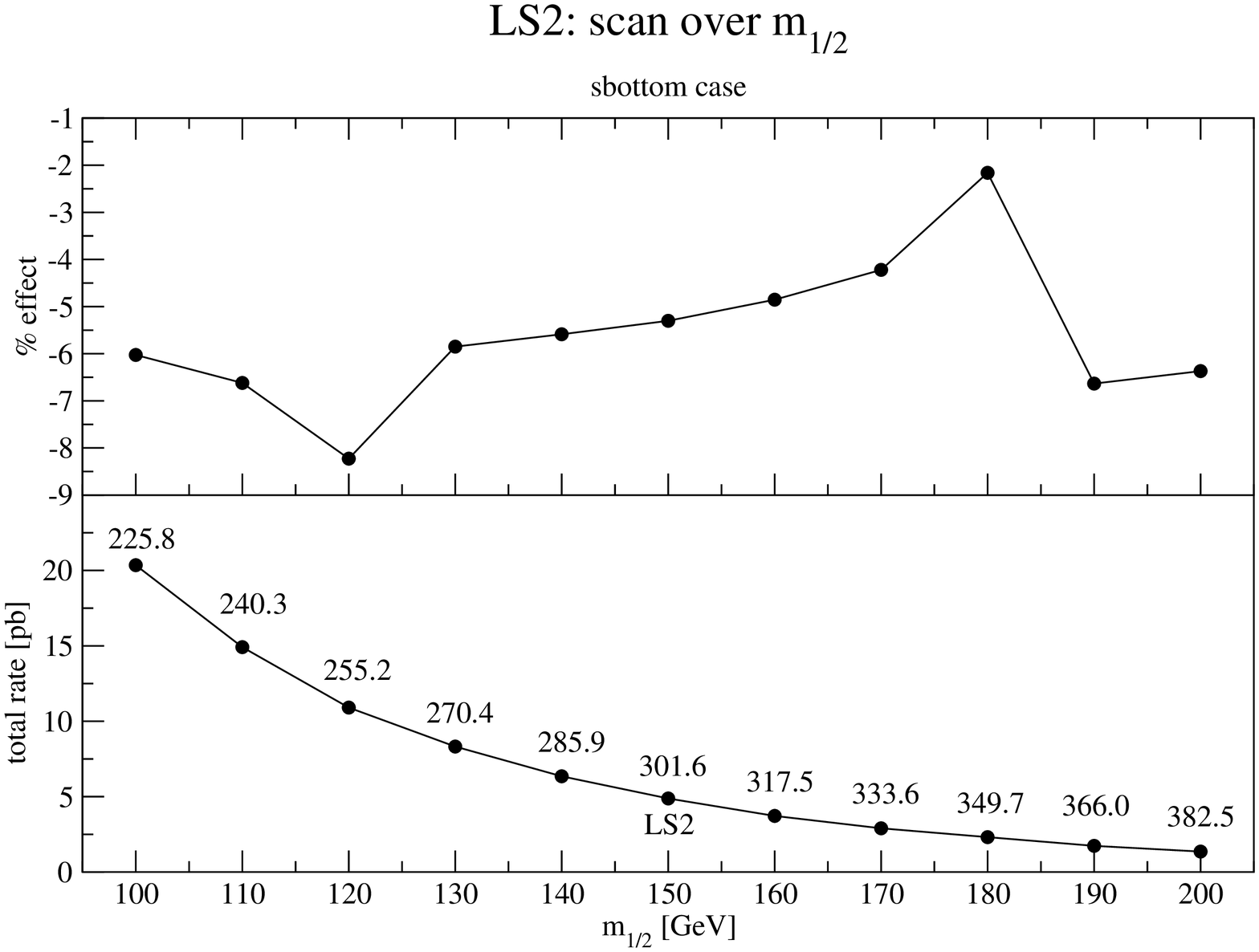, width=0.88\textwidth, angle=0}
\caption{LS1 and LS2: scan over the mSUGRA parameter $m_{1/2}$ for diagonal sbottom production. The top panels show the percentual effect on the integrated cross section, the bottom panels show the variation in the value of the total cross section; the numbers above the curves in the bottom panels represent the value of the sbottom mass $m_{\tilde b_1}$(in GeV).}
\label{fig:m12sbottom}
\end{figure}

\begin{figure}
\epsfig{file=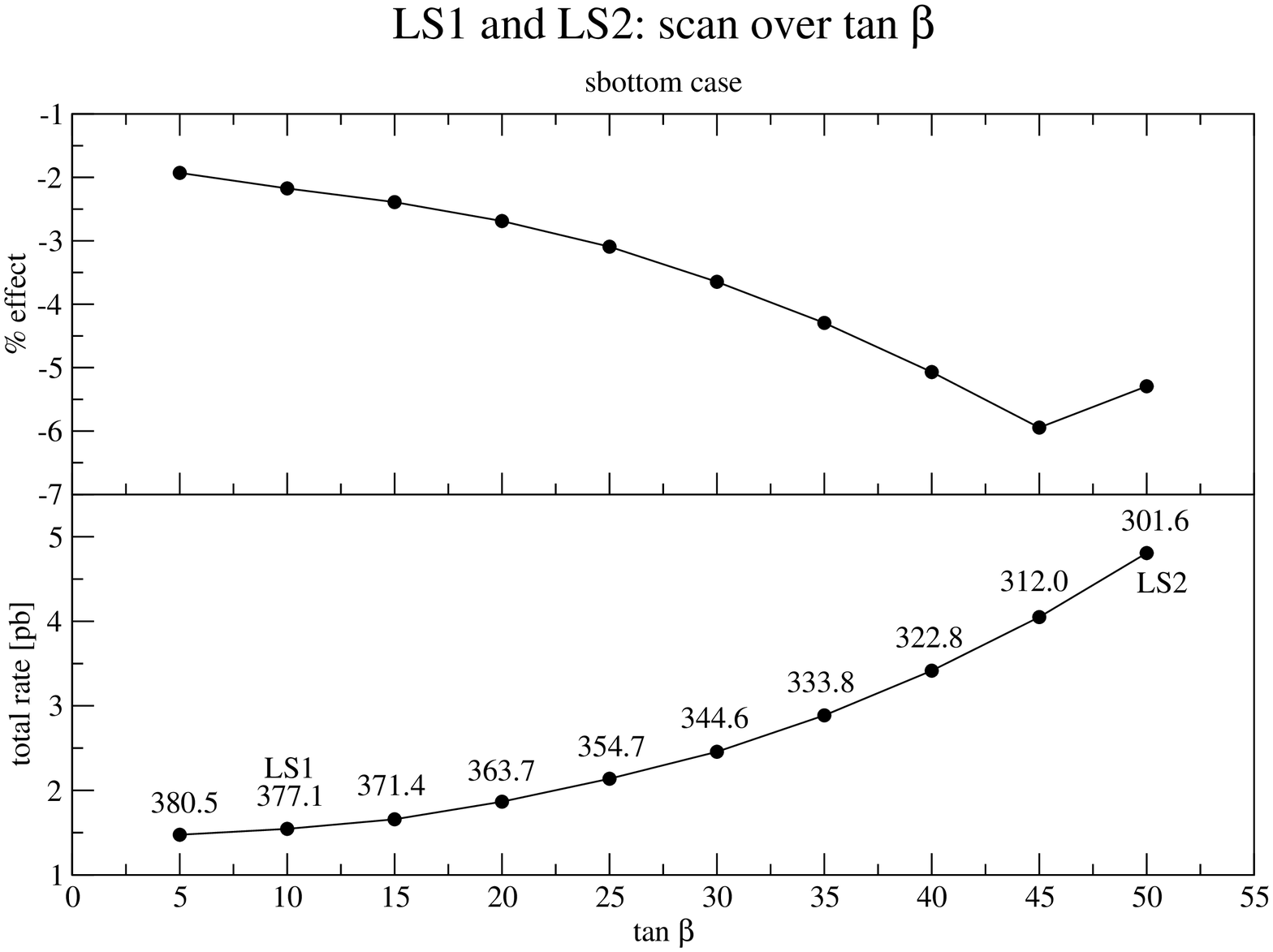, width=0.88\textwidth, angle=0}
\caption{LS1 and LS2: scan over the mSUGRA parameter $\tan\beta$ for diagonal sbottom production. The top panels show the percentual effect on the integrated cross section, the bottom panels show the variation in the value of the total cross section; the numbers above the curves in the bottom panels represent the value of the sbottom mass $m_{\tilde b_1}$(in GeV).}
\label{fig:tanbetasbottom}
\end{figure}

Our search of extra SUSY parameter dependence has been performed in the following way. For each benchmark point, we have varied in turn one of the four conventional parameters ($\tan\beta$, $m_{1/2}$, $m_0$ and $A_0$) in a reasonable range, and computed the variable relative one loop effect and rate. For practical reasons we have only considered in the diagonal stop case the largely dominant $\tilde{t}_1\tilde{t}_1^*$ component, and have shown the value of the $\tilde{t}_1$ mass which is generated by the variation of the chosen parameter. We anticipate, to shorten our presentation, that for diagonal stop production we shall only show in Figs.~(\ref{fig:tanbetam12SPS5}-\ref{fig:tanbetam12SPS1aslope}) the complete numerical results for those cases that seem to us reasonably meaningful, in particular that correspond to a total rate not below the 1~pb (extreme?) limit. This choice selects the set of LS1, LS2, SPS1a, SPS1a', SPS1a slope and SPS5 benchmark points, but to perform a comparison with Ref.\cite{3} we have also included the (perhaps academical) case of SPS2. For the remaining benchmark points, given the negligible value of their rates, we have only shown, for academic information, the dominant relative one loop effects in Fig.(\ref{fig:remainingbenchmarks}). \newline
Figs.(\ref{fig:tanbetam12SPS5},\ref{fig:m0A0SPS5}) show the results that we have obtained for the point SPS5, which is perhaps the most relevant one. As a general feature, common to all the considered cases, one sees that the SUSY one loop effects are almost systematically negative and small, of the few percent size. For what concerns the dependence on the chosen parameter, one sees for SPS5 that the variation of $m_{1/2}$ can produce a maximal variation of the relative effect of approximately three percent. A smaller variation, of approximately 1.5 percent, is generated in the considered range of $\tan\beta$. Varying $m_0$ and $A_0$ has essentially no practical effect ($\sim$ below one percent) on the one loop contribution. The latter remains, in all cases, of the few percent at most.\newline
In Figs.(\ref{fig:tanbetaSPS1SPS2}-\ref{fig:tanbetam12SPS1aslope}), to save space, we have only shown the maximal relative variation and the corresponding parameter. This choice selects $\tan\beta$ for SPS1a, $m_{1/2}$ for SPS1a', $\tan\beta$ for SPS2, $m_{1/2}$ for LS1 and LS2. In the SPS1a' case we have also plotted the $\tan\beta$ dependence to perform a comparison with Ref.\cite{3}. In all cases, the relative one loop effect is negative and small, typically of the one-two percent size.\newline
A special case is that of the benchmark point SPS1a slope, where the parameters $m_{1/2}$, $m_0$ and $A_0$ are related. In this case, we have plotted in Fig.~(\ref{fig:tanbetam12SPS1aslope}) the variations with $m_{1/2}$ and $\tan\beta$. One sees that in the first case the relative negative effect can vary between one and four percent, remaining often in the three-four percent range. This represents the most relevant extra parameter dependence of our stop analysis. Varying $\tan\beta$ can produce a smaller ($\sim$ 1.5) effect, with an overall negative relative contribution in the five percent region which a priori might be visible with a dedicated experimental search. \newline
An important step is now the comparison with previous results. Concerning the total rates, one can see from our curves that the values obtained for the points SPS1a, SPS1a', SPS2 and SPS5 essentially reproduce, taking the corresponding stop mass values, the gluon-gluon component of Ref.\cite{3} Table 1. For the parameter dependence, Ref.\cite{3} shows the SPS1a' case but uses, apart from $\tan\beta$, a different set of parameters. A comparison of the $\tan\beta$ dependences for this point shows a qualitative agreement, i.e. a small and negative effect that increases with $\tan\beta$, although our values are slightly larger, in the three percent range. We conclude in this case, in full agreement with Ref.\cite{3}, that the dependence on the extra SUSY parameters is for SPS1a' extremely small. \newline
Another comparison can be performed for the SPS1a slope and SPS5 cases with the plots of Ref.\cite{2}. Again one can see an essential agreement between our one-lop results and the Born results of Ref.\cite{2}, as one would expect given the smallness of our one loop effects. \newline
The conclusion from our analysis of diagonal stop antistop production is that supersymmetric contributions due to extra SUSY parameters exist, but are generally apparently too small, at the few percent level, to produce an appreciable effect under realistic LHC experimental conditions, at least in a first luminosity phase. Our next step has been that of repeating our analysis for the diagonal sbottom-antisbottom production. Here we have only considered the LS1 and LS2 points, that would have a rate of the pb size. The results of our calculation are shown in the next Figures, that we now briefly comment. \newline
As one sees from Figs.(\ref{fig:m0sbottom}-\ref{fig:tanbetasbottom}), the dependence of the effects on $m_0$ and $A_0$ is essentially negligible. For $m_{1/2}$ there is also no dependence on LS1, and a larger but irregular dependence (same values for different $m_{1/2}$) on LS2. The dependence on $\tan\beta$ exhibits a different, and possibly appreciable, feature. One sees that the negative effect regularly increases with $\tan\beta$, like in the stop cases, but changing more, i.e. from $\sim$2 percent to $\sim$6 percent in the explored range. In particular, we believe that a relative effect of approximately six percent, in correspondence to a rate of the 5 pb size, might be, in principle, proposed for a highly dedicated experimental search.

In conclusion, a reasonable picture that seems to emerge from our combined analysis of the stop-antistop and sbottom-antisbottom diagonal production processes is that of a possible, although mild, dependence of the one-loop electroweak effect in the mSUGRA scenario on extra parameters. Keeping this result in mind, we have also computed, for all benchmark points, the non diagonal total rate derived from one-loop $gg$ electroweak diagrams. We remind the reader that a calculation of the non diagonal rate, derived from $q\overline{q}$ annihilation at Born level via $Z$ exchange, already exists~\cite{2} for SPS5. In the stop case, the value that is obtained is larger than that coming from the kinematically depressed NLO QCD diagrams, and is equal to $\simeq 6\cdot 10^{-4}$~pb. In the sbottom case, the value that is obtained is equal to $\simeq 1.5\cdot 10^{-5}$~pb.
In Tabs.~(\ref{tab:nondiagonalstop},\ref{tab:nondiagonalsbottom}), we show the values that we have derived for the different benchmark points both for the one-loop $gg$ diagrams and from the $Z$ exchange calculations. One sees that the one-loop electroweak values are of the same size as those due to $Z$ exchange and in some cases larger. This could have some relevance for the {\em meaningful} benchmark points. For example, in the LS2 stop production case, summing the one-loop with the $Z$-exchange contribution, one would get a total rate of approximately $10^{-2}$~pb. This is a factor $15$ larger than the SPS5 point of~\cite{2}, but realistically hard for experimental detection. A similar conclusion might be drawn for the rates of the remaining meaningful points if one sums the one-loop with the $Z$-exchange contributions. The results we have obtained for the sbottoms are similar, but because of the tiny cross sections involved, the experimental confirmation of our predictions will be again problematic.

% Nevertheless, we have identified a sort of pattern in the differences between $gg$- and $q\overline{q}$-initiated processes. Both for stop and sbottom production, the cross section is larger in the $gg$ case only for the scenarios LS2, SPS4 and SU6: these points share the same value of $\tan\beta = 50$, so it seems that in the non-diagonal production process this parameter might be able to affect the observables we have considered.
% 
% In this spirit, we show only for the LS2 case in Fig.~(\ref{fig:distrib:nondiag:LS2}) the detailed invariant mass distribution. 

\section{Conclusions}

We have devoted our analysis to the search for extra (i.e. different from the final squark masses) parameters dependence in the processes of diagonal and non-diagonal stop-antistop and sbottom-antisbottom production from the $gg$ initiated channel at EW NLO at LHC. With this aim, we have chosen twelve representative mSUGRA benchmark points and performed a variation of the mSUGRA parameters. We have verified in all cases the presence of a small (at the few percent level) relative difference of the effects with a more important role apparently played by different parameters for different points, in particular by $\tan\beta$ in a case of sbottom production.

% The most acceptable solution appears to us to be that this difference is due to the single parameter $\tan\beta$. The $\tan\beta$ dependence of the effects is quite similar to that exhibited by a negative component of the logarithmic Sudakov expansion, which increases with $\tan\beta$. This could be qualitatively understood since the requested production energies in all processes would become eventually larger than the mass scale M of Eqs.(\ref{sudakov2}),(\ref{sudakov3}) and the corresponding negative effect might remain effective in the total rate. In non-diagonal production, a certain dependence on $\tan\beta$ also seems to exist.\newline

Certainly, the possibility of experimental verification of our conclusions would require very high luminosity scenarios and accuracies, representing a real challenge for the LHC experimental groups. This might, though, become interesting in case of a previous LHC supersymmetric production, which might justify the idea of the dedicated experimental effort that we have mentioned. In this respect, we should mention that the possibility of a determination of SUSY parameters dependence from the process of stop-chargino production has been already considered by us at Born level in a previous paper~\cite{Beccaria:2006wz}. In view of the results obtained in this present search, we are now considering the derivation of the EW one-loop effects on stop-chargino production. The calculation is already in progress.

\newpage

\begin{table}[h]
 \begin{tabular}{|c|ccccc||cc|cc|}
 \hline
 ~mSUGRA scenario~ & $m_0$ & $m_{1/2}$ & $A_0$ & $\tan\beta$ & $\textrm{sign } \mu$ & $\quad m_{\widetilde{t}_1} \quad$ & $\quad m_{\widetilde{t}_2}\quad$ & $\quad m_{\widetilde{b}_1}\quad$ & $\quad m_{\widetilde{b}_2}\quad$ \\
 \hline
 LS1         & 300           & 150 & -500           & 10 & + & 214.6 & 460.5 & 377.1 & 444.7 \\
 LS2         & 300           & 150 & -500           & 50 & + & 224.7 & 430.4 & 301.6 & 399.3 \\
 SPS1a       & 100           & 250 & -100           & 10 & + & 399.7 & 585.5 & 515.7 & 546.6 \\
 SPS1a'      & 70            & 250 & -300           & 10 & + & 367.3 & 581.9 & 504.4 & 541.7 \\
 SPS1a slope & ~0.4$m_{1/2}$ & 250 & -0.4$m_{1/2}$  & 10 & + & 399.7 & 585.5 & 515.7 & 546.6 \\
 SPS2        & 1450          & 300 & 0              & 10 & + & 921.4 & 1289  & 1279  & 1540  \\
 SPS3        & 90            & 400 & 0              & 10 & + & 645.2 & 840.3 & 790.1 & 823.7 \\
 SPS4        & 400           & 300 & 0              & 50 & + & 540.1 & 692.5 & 614.9 & 687.2 \\
 SPS5        & 150           & 300 & -1000          & 5  & + & 279.0 & 651.2 & 566.3 & 651.1 \\
 SPS6        & 150           & 300 & 0              & 10 & + & 494.6 & 675.6 & 617.0 & 649.4 \\
 SU1         & 70            & 350 & 0              & 10 & + & 566.4 & 754.0 & 698.6 & 729.8 \\
 SU6         & 320           & 375 & 0              & 50 & + & 634.1 & 794.7 & 712.1 & 785.8 \\
 \hline  
 \end{tabular}
\caption{mSUGRA benchmark points and masses of stops and sbottoms (all the values are in GeV)}
\label{tab:bench}
\end{table}

\begin{table}
 \begin{tabular}{|c|r@{ @ }l|r@{ @ }l|}
 \hline
 {} & \multicolumn{2}{|c|}{$\sigma_{g g \to \begin{small}\tilde{t}_1\tilde{t}_1^*\end{small}}$} & \multicolumn{2}{|c|}{$\sigma_{g g \to \tilde{b}_1\tilde{b}_1^*}$} \\
 \hline
 LS1      & ~27.00   & $m_{\tilde t_1} = 214.6$~GeV~ & 1.54     & $m_{\tilde b_1} = 377.1$~GeV~ \\
 LS2      & 21.51    & $m_{\tilde t_1} = 224.7$~GeV~ & 4.85     & $m_{\tilde b_1} = 301.6$~GeV~ \\
 SPS5     & 7.46     & $m_{\tilde t_1} = 279.0$~GeV~ & 0.156    & $m_{\tilde b_1} = 566.3$~GeV~ \\
 ~SPS1a'~ & 1.76     & $m_{\tilde t_1} = 367.3$~GeV~ & 0.30     & $m_{\tilde b_1} = 504.4$~GeV~ \\
 SPS1a    & 1.10     & $m_{\tilde t_1} = 399.8$~GeV~ & 0.261    & $m_{\tilde b_1} = 515.7$~GeV~ \\
 SPS6     & 0.33     & $m_{\tilde t_1} = 494.6$~GeV~ & 0.0908   & $m_{\tilde b_1} = 617.0$~GeV~ \\
 SPS4     & 0.19     & $m_{\tilde t_1} = 540.1$~GeV~ & 0.090    & $m_{\tilde b_1} = 614.9$~GeV~ \\
 SU1      & 0.147    & $m_{\tilde t_1} = 566.4$~GeV~ & 0.0416   & $m_{\tilde b_1} = 698.6$~GeV~ \\
 SU6      & 0.073    & $m_{\tilde t_1} = 634.1$~GeV~ & 0.0358   & $m_{\tilde b_1} = 712.1$~GeV~ \\
 SPS3     & 0.066    & $m_{\tilde t_1} = 645.2$~GeV~ & 0.0185   & $m_{\tilde b_1} = 790.1$~GeV~ \\
 SPS2     & ~0.00617 & $m_{\tilde t_1} = 921.4$~GeV~ & ~0.00052 & $m_{\tilde b_1} = 1279$~GeV~ \\
 \hline
 \end{tabular}
\caption{Total cross-sections (in pb) for diagonal stop and sbottom production. The point SPS1a slope has not been included since it coincides with SPS1a at $m_{1/2} = 250$~GeV.}
\label{tab:diagonal}
\end{table}

\begin{table}
\begin{tabular}{|c|c|c|}
\hline
{} & $\sigma_{q \bar q \to \tilde{t}_1\tilde{t}_2^*+\tilde{t}_2\tilde{t}_1^*}$ & $\sigma_{g g \to \tilde{t}_1\tilde{t}_2^*+\tilde{t}_2\tilde{t}_1^*}$ \\
\hline
LS2 & 0.0034 & 0.0058 \\
LS1 & 0.0026 & 0.0012 \\
SPS5 & 0.00057 & 0.00049 \\
~SPS1a~ & 0.00054 & 0.00038 \\
SPS6 & 0.00022 & 0.00013 \\
SPS4 & 0.00017 & 0.00045 \\
SU1 & 0.00011 & 0.000057 \\
SU6 & 0.000080 & 0.00016 \\
SPS3 & 0.000057 & 0.000024 \\
~SPS2~ & 0.00000044 & 0.00000023\\
\hline
\end{tabular}
\caption{Total cross-sections (in pb) for non-diagonal stop production starting from $q\bar q$ and $g g$.}
\label{tab:nondiagonalstop}
\end{table} 

\begin{table}
\begin{tabular}{|c|c|c|}
\hline
{} & $\sigma_{q \bar q \to \tilde{b}_1\tilde{b}_2^*+\tilde{b}_2\tilde{b}_1^*}$ & $\sigma_{g g \to \tilde{b}_1\tilde{b}_2^*+\tilde{b}_2\tilde{b}_1^*}$ \\
\hline
LS2 & 0.0027 & 0.011 \\
LS1 & 0.00020 & 0.000024 \\
SPS5 & 0.000013 & 0.00000087 \\
~SPS1a~ & 0.00020 & 0.000049 \\
SPS6 & 0.000068 & 0.0000067 \\
SPS4 & 0.00016 & 0.0006 \\
SU1 & 0.000040 & 0.0000032 \\
SU6 & 0.000081 & 0.00024 \\
SPS3 & 0.000017 & 0.0000012 \\
~SPS2~ & $2.49\times10^{-9}$ & $2.1\times10^{-10}$ \\
\hline
\end{tabular}
\caption{Total cross-sections (in pb) for non-diagonal sbottom production starting from $q\bar q$ and $g g$.}
\label{tab:nondiagonalsbottom}
\end{table}

\newpage
\hfill

\hfill

\hfill

\hfill

\hfill

\hfill

\hfill

\hfill

\hfill

\hfill

\hfill

\hfill

% -----------------------------------------------------

\clearpage

\end{document}